\documentclass[12pt]{article}
\usepackage[a4paper,tmargin=2cm,bmargin=2cm,lmargin=2cm,rmargin=2cm]{geometry}
\usepackage[T1]{fontenc}
\usepackage[utf8]{inputenc}
\usepackage{amssymb}
\usepackage{amsmath}
\usepackage{times}
\usepackage{xcolor}
\usepackage{graphicx}

\usepackage[english]{babel}
\usepackage[colorlinks=true,linkcolor=blue,urlcolor=blue,citecolor=blue]{hyperref}

\newlength{\highlightlength}
\newlength{\highlightlengthbis}

\newcommand{\bea}{\begin{eqnarray}}
\newcommand{\eea}{\end{eqnarray}}
\newcommand{\be}{\begin{equation}}
\newcommand{\ee}{\end{equation}}
\newcommand{\Eq}[1]{Eq.~(\ref{#1})}
\newcommand{\Eqs}[1]{Eqs.~(\ref{#1})}
\newcommand{\eq}[1]{(\ref{#1})}
\newcommand{\rme}{\mathrm{e}}
\newcommand{\rmd}{\mathrm{d}}
\newcommand{\nn}{\nonumber}
\newcommand{\ca}[1]{\mathcal{#1}}

\newcommand{\fig}[2]{\includegraphics[width=#1\columnwidth]{./#2}}

\renewcommand{\a}{\alpha}
\renewcommand{\b}{\beta}
\newcommand{\half}{\frac12}

\usepackage{tikz}
\usetikzlibrary{arrows,shapes,positioning}
\usetikzlibrary{decorations.markings}
\tikzstyle arrowstyle=[scale=1]
\tikzstyle directed=[postaction={decorate,decoration={markings,
    mark=at position .65 with {\arrow[arrowstyle]{stealth}}}}]
\tikzstyle endreversedirected=[postaction={decorate,decoration={markings,
    mark=at position 1.0 with {\arrow[arrowstyle]{stealth}}}}]
\tikzstyle enddirected=[postaction={decorate,decoration={markings,
    mark=at position 1.0 with {\arrow[arrowstyle]{stealth}}}}]
\tikzstyle reverse directed=[postaction={decorate,decoration={markings,
    mark=at position .65 with {\arrowreversed[arrowstyle]{stealth};}}}]

\newcommand{\propdiag}{{{\parbox{1.cm}{{\begin{tikzpicture}[scale=1]
\coordinate (x1) at (0,0) ;
\coordinate (x2) at  (0.6,0) ;
\coordinate (x1p) at  (-.2,0) ;
\coordinate (x2p) at  (0.8,0) ;
\fill (x1) circle (1.5pt);
\fill (x2) circle (1.5pt);
\draw (.3,0) circle (3mm);
\draw [black] (x1) -- (x1p);
\draw [black] (x2) -- (x2p);
\end{tikzpicture}}}}}}

\begin{document}

\title{\bf\Large The two upper critical dimensions of the Ising and Potts models}
\author{\bf\normalsize Kay J\"org Wiese{$^{1}$} and Jesper Lykke Jacobsen{$^{1,2,3}$}}
\date{\small {$^{1}$}CNRS-Laboratoire de Physique de l'Ecole Normale Sup\'erieure, PSL Research University, Sorbonne Universit\'e, Universit\'e Paris Cit\'e, 24 rue Lhomond, 75005 Paris, France.\\
{$^{2}$}Sorbonne Universit\'e, Ecole Normale Sup\'erieure, CNRS, Laboratoire de Physique (LPENS).\\
{$^{3}$}Institut de Physique Th\'eorique, CEA, CNRS, Universit\'e Paris-Saclay.}

\maketitle

\begin{abstract}
We derive the  exact  actions of the $Q$-state Potts model valid on any graph, first for the spin degrees of freedom, and second for the   Fortuin-Kasteleyn clusters. In both cases the  field   is a  traceless $Q$-component scalar field  $\Phi^\alpha$.  
For the  Ising model ($Q=2$),   the field theory for the spins has upper critical dimension $d_{\rm c}^{\rm spin}=4$, whereas  for the clusters it has $d_{\rm c}^{\rm cluster}=6$.
As a consequence, the probability for three points to be in the same cluster is   not given by  mean-field theory   for $d$ within $4<d<6$. We estimate the associated universal structure constant as $C=\sqrt{6-d}+ \ca O(6-d)^{3/2}$.  This shows that some observables in the Ising model have an upper critical dimension of 4, while others have an upper critical dimension of $6$. 
Combining perturbative results from the $\epsilon=6-d$ expansion with a non-perturbative treatment close to dimension $d=4$ allows us to locate the shape of the critical domain of the Potts model in the whole $(Q,d)$ plane.

\end{abstract}

\section{Introduction}
\begin{figure}[t]
\centerline{\fig{0.65}{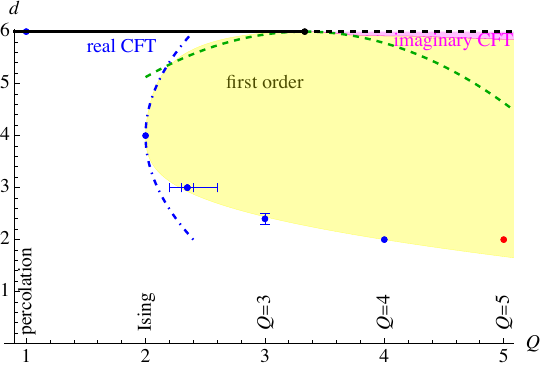}}
\caption{The various critical dimensions present in the Potts model, as explained in the main text. The yellow shaded region marked as ``first order'' is the area without a real critical CFT. 
The upper boundary is obtained via NPRG/Wilson. The left upper branch for $Q<Q_{\rm c}$ and $d>4$ is numerically very robust. The upper right branch for $Q>Q_{\rm c}$ and $d>4$ is delicate, and may change (probably further reduce) in higher orders. The lower bound has been studied in the literature; here we   use the polynomial fit from \Eq{curvatureNPRG}, with two additional powers added, adjusted such that $Q_{\rm c}(d=6)=10/3$ and $Q_{\rm c}(d=4)=4$.
Points on the boundary are critical. The point $Q=5$ in $d=2$ has complex couplings. The domain denoted as ``imaginary CFT'' (in pink) represents a critical theory if all couplings are real after rotation of $\phi\to i \phi$, i.e.\ odd couplings purely imaginary, and even couplings real. 
For $Q\to \infty$, perturbatively it gives $Q$ decoupled Lee-Yang theories (see section \ref{s:upper-right-branch} for a caveat). If one starts with real couplings, as in a simulation,   this region is   first order.}
\label{f:Potts-phase-diagram}
\end{figure} 
The Ising and $Q$-state Potts models have a long history \cite{Wu1982}. Let us denote by $\vec s_x$ the state variable,   allowed to take $Q$ different values. The energy of the Potts-model on a graph $\ca G=(\ca V, \ca E)$ with vertices $\ca V$ and edges $\ca E$ is defined by 
\bea
\label{H-Potts}
{\cal H}^{\rm Potts}_Q[\vec s;\vec h] = -\sum _{(x,y)\in \ca E} J \delta_{\vec s_x, \vec s_y}-    \sum_{x\in \ca V} \vec h_x    \vec s_x, 
\eea
where the first term is $-J$ if $\vec s_x$ and $\vec s_y$ are in the same state and zero otherwise. The last term evaluates to $h_\alpha$ if the   spin is in state $\a$. 
The Ising model is the special case with  $Q=2$.
While the Potts model is originally defined for integer $Q$, it can be extended to any $Q\in \mathbb C$ via the Fortuin-Kasteleyn cluster expansion \cite{FortuinKasteleyn1972}, see section \ref{s:Fortuin-Kasteleyn}. A natural question to ask is whether 
these two expansions lead to the same field theory. 
While for the spin degrees of freedom, the latter  was derived in the classical work by Golner \cite{Golner1973}, Zia and Wallace  \cite{ZiaWallace1975},  Amit \cite{Amit1976}, and Priest and Lubensky \cite{PriestLubensky1976b,PriestLubensky1976}, a field theory for the cluster expansion is lacking.
This comes   with the pressing question of whether the leading non-Gaussian term  is cubic as in  \cite{Amit1976,PriestLubensky1976b,PriestLubensky1976,AlcantaraBonfirmKirkhamMcKane1981}, or quartic as in \cite{ZiaWallace1975}.
This   is  relevant as the upper critical dimension is six for a cubic interaction, and   four for a quartic one. 

On Fig.~\ref{f:Potts-phase-diagram} we show   critical  
values of $Q_{\rm c}(d)$ beyond which the critical point disappears, or equivalently $d_{\rm c}(Q)$ beyond which this happens:
Values found in the  literature, are $d_{\rm c}(Q=1)=6$ for percolation
\cite{Amit1976,Wu1978,PriestLubensky1976b,PriestLubensky1976}, $d_{\rm c}(Q=2)=4$ for the Ising model \cite{WilsonFisher1972}, $d_{\rm c}(Q=3)\approx 2.5$ from the numerical conformal bootstrap \cite{ChesterSu2022}, and $d_{\rm c}(Q=4)=2$ \cite{QianDengLiuGuoBlote2016,Hartmann2005}. 
The situation in $d=3$ seems debated, with values for the critical value of $Q$ ranging from $2.2$ to $2.6$: $Q_{\rm c} \approx 2.2$ via an
Ornstein-Zernicke approximation 
\cite{GrollauRosinbergTarjus2001}, 
$Q_{\rm c}=2.35(5)$ via MC
\cite{Hartmann2005},   $Q_{\rm c}=2.2$ via real-space RG \cite{NienhuisRiedelSchick1981}, 
$Q_{\rm c}=2.11(7)$ via NPRG \cite{Sanchez-VillalobosDelamotteWschebor2023}, $Q_{\rm c}=2.1$ as well as $Q_{\rm c}=2.45(1)$  \cite{LeeKosterlitz1991} and $Q_{\rm c} = 2.620(5)$ \cite{Gliozzi2002}, both via MC.
The most precise value seems to be from Ref.~\cite{Hartmann2005}.
In $d=2$ the critical value is $Q_{\rm c}(d=2)=4$ \cite{Baxter1973,MazzarisiCorberiCugliandoloPicco2020,Duminil-CopinSidoraviciusTassion2017}.

The green dashed curve is our  result derived in section \ref{The upper boundary of the non-critical domain}, 
\be\label{dc-6-eps}
d_{\rm c} = 6-\frac{729}{1480}(Q-Q_{\rm c})^2 + \ca O(Q-Q_{\rm c})^3.
\ee
In blue dot-dashed is shown the 
result in the LPA${}'$-approximation of the NPRG \cite{Delamotte2012,DupuisCanetEichhornMetznerPawlowskiTissierWschebor2021},   used for Potts in \cite{ZinatiCodello2018,Sanchez-VillalobosDelamotteWschebor2023}. In this scheme,
  we find in agreement with \cite{Sanchez-VillalobosDelamotteWschebor2023}
\be\label{curvatureNPRG}
Q_{\rm c}-2 \approx 0.10 (4-d)^2 + \ca O(4-d)^3.
\ee
There is an older estimation  \cite{NewmanRiedelMuto1984}, obtained via the numerical solution of eleven coupled RG equations in the Wilson scheme, which reads
\be\label{NewmanRiedelMuto1984}
Q_{\rm c}-2 \approx 0.153 (4-d)^2 + \ca O(4-d)^3.
\ee
We could not reproduce this result, see section \ref{s:Non-perturbative renormalization}.
Finally, 
Refs.~\cite{AharonyPytte1981,NewmanRiedelMuto1984}   state that $Q_{\rm c}=2$ for $d>4$.  This is incompatible with the expansion \eq{dc-6-eps}. 
On the other hand, both  expansions    make sense if we use 
\Eq{NewmanRiedelMuto1984} or  \eq{curvatureNPRG} also in dimension $d>4$: As  a glance at Fig.~\ref{f:Potts-phase-diagram} shows,  
the expansions in \Eqs{dc-6-eps} and \eq{NewmanRiedelMuto1984} or \eq{dc-6-eps}  and \eq{curvatureNPRG} are compatible, and taken together allow for a rather precise delimination of the critical region for $d>4$ and $Q\le 10/3$. This can be  obtained within   various non-perturbative renormalization schemes (see section \ref{s:Non-perturbative renormalization}). 
For $Q>10/3$ and $d\to 6$ a region (pink in Fig.~\ref{f:Potts-phase-diagram}) appears with a purely imaginary coupling. As we discuss in section \ref{The upper boundary of the non-critical domain}, this may be an artifact of the expansion. If we start with real couplings,  the yellow first-order region   covers this region as well.

In the  argumentation above we    equated an RG flow to strong coupling with a first-order transition, as is implicitly (i.e.\ without proof) commonly done in the literature. 
We cannot exclude that the flow which apparently goes to strong coupling is towards a critical fixed point not accessible in  any of    our schemes. 
In this scenario, the boundary of the critical regime remains unchanged, but the interior of the ``first-order'' region may become second-order, or split in a first-order and a second-order regime.  
 
To complement our introduction,  let us mention    field-theory  results for the cubic   \cite{AlcantaraBonfirmKirkhamMcKane1981,SafariVaccaZanusso2020,BorinskyGraceyKompanietsSchnetz2021,KompanietsPikelner2021},   quartic 
 \cite{SafariVaccaZanusso2020} and quintic theories \cite{CodelloSafariVaccaZanusso2020}, which each are consistent  below their respective upper critical dimensions, and which we expect to be (higher critical) conformal field theories (CFT). 
Expansions as in \Eqs{NewmanRiedelMuto1984} or \eq{curvatureNPRG} also appear in  \cite{NahumPhD}.

\medskip

In this work, we first derive  the exact field-theoretical action in the spin formulation (section \ref{s:spin-expansion}),  followed by the exact   action of the cluster expansion (section \ref{s:Fortuin-Kasteleyn}).
By {\em exact} we mean that on an arbitrary graph   the field theory   produces the partition function and all correlation functions, without any approximation. 

When evaluating these exact actions for regular lattices,   e.g.\ the cubic lattice in $d$ dimensions, and taking a continuum limit, 
the action   expanded   in the fields contains an infinity of interactions. In order to set up an efficient RG scheme, we follow the standard procedure to retain  solely the most relevant terms. The result of this procedure is different in the spin and cluster expansions: 
In the cluster expansion, the leading interaction is  cubic  for all $Q$, leading to $d_{\rm c}^{\rm cluster}(Q)=6$. 
In the spin formulation, the Ising model is special, since its symmetry  under a reversal of the field (the magnetization) excludes a cubic vertex, leading to an upper critical dimension  $d_{\rm c}^{\rm spin}(Q=2)=4$. 
This suggests that for the Ising model spin-correlation functions are given by mean-field theory in dimensions between 
four and six. However, the probability that three sites are   in the same cluster is given by the non-trivial cubic theory. The latter predicts  to leading order in $6-d$ that 
\be\label{C}
 C:= \frac{\ca P(x,y,z)}{\sqrt{\ca P(x,y) \ca P(y,z), \ca P(z,x)}} = \sqrt{ 6-d} + \ca O (6-d)^{3/2} ,
\ee
where $\ca  P(x,y,z)$ is the probability that $x$, $y$, and $z$ are in the same cluster, and likewise for the terms in the denominator. While the functional form of ${\ca P(x,y,z)}$ is imposed by global conformal invariance \cite{DiFrancescoMathieuSenechal} as written, the amplitude is non-trivial, and measurable in a numerical simulation.  
We report below in \Eq{C-RG} a similar result for other values of $Q$. 

The article is organized as follows: 
In section \ref{s:spin-expansion}, we derive the field theory for the spin expansion, followed by 
the field theory for the cluster expansion in section \ref{s:Fortuin-Kasteleyn}.
Section \ref{s:Renormalization: 6-epsilon expansion} treats the ensuing field theories perturbatively, 
and calculates the structure constant. 
In section \ref{s:Non-perturbative renormalization} we treat the $Q$-state Potts model 
via the non-perturbative renormalization group, which allows us to draw the phase diagram in the 
whole $(Q,d)$ plane. 
In section \ref{s:Conclusion} we   conclude. 
Some technical details on the Potts algebra are relegated to appendix \ref{a:algebraic-objects}.

\section{Spin expansion}
\label{s:spin-expansion}
We start with a reminder or the spin expansion for the Potts model. 
Consider the $Q$-state Potts model. Following \cite{Golner1973,ZiaWallace1975}, we represent each state $\alpha=1,...,Q$ as a vector\footnote{While this representation is heuristically appealing, and allows one to  coarse-grain spin degrees of freedom, it may be but one of various distinct possibilities. We will see a   systematic procedure for the cluster expansion in section \ref{s:Fortuin-Kasteleyn}.} $$\vec s_x\in \{\vec e_1, ...,\vec e_{Q}\},  
$$  of length $\sqrt{1-\frac1Q}$, and a scalar-product of $-1/Q$ between distinct vectors,
\bea
\label{algebra-1}
\sum _\alpha  \vec e_\a &=&0 \\
\label{algebra-2}
 \vec e_\a \cdot \vec e_\beta &:=& 
\sum_{i}  e_\alpha^i    e_\beta^i =   \delta_{\alpha\beta} - \frac1Q \\
\label{algebra-3}
  e^i \circ e^j &:=& \sum_{\alpha} e^i_\alpha  e^j_\alpha =   \delta ^{ij}  .
\eea
The normalizations are chosen for convenience. 
As an example, for $Q=2$ we have $\vec e_1 = -\vec e_2= 1/\sqrt2$, while for $Q=3$ the three vectors lie in the plane with an angle of $2\pi/3$ between them. 
We refer to appendix \ref{a:algebraic-objects} for details on this construction. 

The energy of the Potts-model in this spin representation, in the presence of a magnetic field $\vec h_x$  is 
\be\label{H-Potts1}
{\cal H}[\vec s;\vec h] = -\sum _{(x,y)\in \ca E} J \delta_{\vec s_x, \vec s_y}-    \sum_{x\in \ca V} \vec h_x    \vec s_x.
\ee
According to \Eq{algebra-2}, 
\be
\vec s_x\cdot \vec s_y +\frac 1 Q =  
\begin{cases}
1 & \mbox{ if } \vec s_x= \vec s_y \\
0 & \mbox{ else}
\end{cases}.
\ee
With the help of this identity, 
\Eq{H-Potts1} can be rewritten as 
\bea \label{H-Potts2}
{\cal H}[\vec s;\vec h] &=& - \sum_{(x,y)\in \ca E}  {J}  \left[ \vec s_x\cdot \vec s_y+\frac 1Q\right] -     \sum_{x\in \ca V} \vec h_x    \vec s_x~~~~ \nn\\
&=& \frac{J}2 \sum_{(x,y)\in \ca E}\Big\{      \left[ \vec s_x- \vec s_y\right]^2 -  2 \Big\}  -    \sum_{x\in \ca V} \vec h_x    \vec s_x.~~~~
\eea
To obtain the effective action, we follow the standard procedure \cite{Amit}: introduce an auxiliary field to decouple the interaction, sum over the spins, and finally perform a Legendre transform w.r.t.\ the auxiliary field. 
We start with 
\be
\ca Z[\vec h]\equiv \rme^{-\ca W[\vec h]}=\left< \rme^{-\half\sum_{{x,y \in \ca V}} \vec s_x K_{xy} \vec s_y+\sum_{x\in \ca V}\vec h_x \vec s_x} \right>_{\!\!\vec s}.
\ee
As $\vec s_x^2=1-\frac1Q$, the matrix element  $\ca K_{xx}$ can be chosen to our liking, allowing us to ensure that the inverse kerne $\ca K_{xy}^{-1}$ exists. The double sum over $x,y$   runs over all vertices, $x\in \ca V$ and $y\in \ca V$, but $K_{xy}=0$ if $(x,y) \not\in \ca E$. To reduce clutter, we drop the notation that $x,y \in \ca V$ for the remainder of this section.

We now decouple the interaction, 
\bea
\rme^{-\ca W[\vec h]}&=&\left< \rme^{-\half\sum_{{x,y}} \vec s_x K_{xy} \vec s_y+\sum_{x}\vec h_x \vec s_x} \right>_{\!\!\vec s}\nn\\
&=& {\cal N} \int \prod_{x} \rmd \phi_x \,\rme^{{\half\sum_{x,y} \vec \phi_x K^{-1}_{xy} \vec \phi_y }}
\left< \rme^{ \sum_{x}( \vec \phi_x+{\vec h_x)  \vec s_x }} 
\right>_{\!\!\vec s}.
\eea
Note that in order for the path integral to converge,  $\vec \phi$ is chosen imaginary. Summing over spins  yields
(for each site $x$, dropping the index)
\bea\label{V}
\sum_{{\vec s \in \{\vec e_{\alpha} \}}}\rme^{( \vec \phi+\vec h)\vec s}&=& \rme^{-V( \vec \phi+\vec h)} ,\\
 V(\vec \phi) &=& -\ln \left( \sum_{n=0}^{\infty} \frac1{n!}\sum_{{\vec s \in \{\vec e_{\alpha} \}}}(  \vec s \cdot \vec \phi)^{n} \right) =-\ln\left( \sum_{n=0}^{\infty} \frac1{n!}\sum_{\alpha=1}^{Q}(  \phi_{\alpha})^{n} \right), 
\eea
where we  defined
\be\label{V2}
\phi_\alpha := \sum _i \phi^i e^i_\alpha \equiv \vec \phi \cdot \vec e_\alpha .
\ee
By construction $\sum_\alpha \phi_\alpha=0$.
This allows us to write
\bea
\rme^{-\ca W[\vec h]}&=& {\cal N}\int \prod_{x} \rmd \phi_x \rme^{{\half\sum_{x,y} \vec \phi_x K^{-1}_{xy} \vec \phi_y }}
\rme^{-V\big(\vec \phi_x+\vec h_x\big)}
\nn\\
&=& {\cal N} \int\prod_{x} \rmd \phi_x \rme^{{\half\sum_{x,y} [\vec \phi_x-\vec h_x] K^{-1}_{xy} [\vec \phi_y-\vec h_x] }}
\rme^{-V\big(\vec \phi_x \big)}\nn\\
&=&{\cal N}' \rme^{{\half }\sum_{x,y} {\vec h}_{x}  K^{{-1}}_{xy} {\vec h}_{y}}  \int \!\prod_{x}\! \rmd \psi_x \rme^{\half\sum_{x,y}  \vec \psi_x  K_{xy}  \vec \psi_y -\vec\psi_x\vec h_y -\sum_{x} {V}\big(\sum_{y} K_{xy} \vec\psi_y\big)}.
\eea
From the first to the second line we  
shifted $\vec \phi_x \to \vec\phi_x-\vec h_x$, and finally replaced $\vec \phi_x$ by $\vec \psi_x := \sum_y K_{xy}^{-1} \vec\phi_y$. The change in measure is reflected by a new normalization constant $\ca N'$.
The last line is 
\be
\rme^{-\ca W[\vec h]}= {\ca N}'  \rme^{{\half }\sum_{x,y} \vec h_x  K^{{-1}}_{xy} \vec h_y} \left< \rme^{\sum_x \vec h_x \vec \psi_x}\right>_{\ca H}
\ee
with the action
(signs: $\rme^{-\ca S[\psi]}$)
\be
\ca S[\vec \psi]= -\half\sum_{x,y}  \vec \psi_x  K_{xy}  \vec \psi_y +\sum_{x} \half {U}(\sum_{y} K_{xy} \vec\psi_{y}).
\ee
We note that ${V}(\vec \phi=\vec 0)=0$ and $\nabla_\phi {V}(\vec\phi)|_{\vec \phi=0} = 0$.
We can therefore write 
\be
 {U(\vec \phi)} =  {\lambda_2}  \vec \phi^2 +     {\lambda_3}  \sum_\alpha (\phi^\alpha)^3 + \lambda_4    \sum_\alpha (\phi^\alpha)^4 + \lambda_{2,2}  \left(\sum_\alpha (\phi^\alpha)^2 \right)^2 + ...
\ee
We   now   construct $\Gamma[\vec \phi]$, the Legendre transform of $\ca W[\vec h]$.
The result, keeping only the most relevant terms, is 
\bea
\Gamma[\vec \phi] &=& \half\sum_{{x,y}}\sum_{\alpha}   \phi_x^\alpha \left[ K_{xy} + \lambda_2 \delta_{xy}\right]   \phi_y^\alpha \nn\\
&& + \half \sum_x \left[  {\lambda_3}  \sum_\alpha (\phi^\alpha_x)^3  + \lambda_4    \sum_\alpha (\phi^\alpha_x)^4 + \lambda_{2,2}  \left(\sum_\alpha (\phi^\alpha_x)^2 \right)^2 + ... \right]\nn\\
&& + \mbox {loop corrections}.
\eea

\section{Fortuin-Kasteleyn  cluster expansion}
\label{s:Fortuin-Kasteleyn}
\subsection{Basics of the cluster expansion}
The Fortuin-Kasteleyn (FK) cluster expansion for the  Hamiltonian \eq{H-Potts1} is \cite{FortuinKasteleyn1972}
\bea
\ca Z[\vec h] &=& \sum_{\{ \vec s_x\}}\rme^{J \sum_{( x,y)\in \ca E}  \delta_{\vec s_x \vec s_y}+\sum_{x\in \ca V} \vec h_x \vec s_x } \nn\\
&=& \sum_{\{ \vec s_x\}}    \prod_{( x,y)\in \ca E} \left[ 1+ (\rme^J {-}1)\delta_{\vec s_x \vec s_y}\right] \rme^{\sum_{x\in \ca V} \vec h_x \vec s_x } \nn\\
&=& \sum_{\{ \vec s_x\}}  \sum_{\ca C}  \prod_{( x,y)\in \ca C}   (\rme^J {-}1)\delta_{\vec s_x \vec s_y}  \rme^{\sum_{x\in \ca V} \vec h_x \vec s_x },
\label{35}
\eea
where $\ca C$ runs over all possibilities to use the term $\delta_{\vec s_x \vec s_y}$, i.e.\ over all subsets of edges $(x,y) \in \ca E$.
Each cluster $\ca C$ is the disjoint union of connected components  $ C_i$, 
\be
\ca C= \dot \cup_i \,{  C}_i.
\ee
We   now interchange the two sums \cite{FortuinKasteleyn1972}, 
\bea
\ca Z[\vec h] &=&  \sum_{\ca C} \sum_{\{ \vec s_x\}}   \prod_{( x,y)\in \ca C}   (\rme^J {-}1)\delta_{\vec s_x \vec s_y}  \rme^{\sum_{x\in \ca V} \vec h_x \vec s_x }
\nn\\
&=& \sum_{\ca C} (\rme^{J}-1)^{|\ca C|} \prod_{ C_i| \dot \cup_i {\ca C}_i= \ca C } \sum_{\vec s} \rme^{\sum_{x \in  C_i} \vec h_x \vec s}.
\eea
The state $\vec s_x= \vec s$ is constant on each connected component $ C_i$, and is independent on two different components. 
If $\vec h_x = \vec 0$, then 
\be
 \sum_{\vec s} \rme^{\sum_{x \in  C_i} \vec h_x \vec s} = Q.
\ee
If $\vec h_x = (h,0,...)$ independent of $x$, then  
\bea
 \sum_{\vec s} \rme^{\sum_{x \in  C_i} \vec h_x \vec s} &=& Q + \rme^{h | C_i|}-1 \nn\\
 &\approx& Q \,\rme^{\frac h Q | C_i|+ \frac{h^2(Q-1)}{2 Q^2} | C_i|^2+ \ca O(h^3)}.
\eea
Thus for $\vec h_x = (h,0,...)$ and $h$ small
\bea
\ca Z[\vec h] &=&\sum_{\ca C} (\rme^{J}-1)^{|\ca C|} Q^{||\ca C||} \rme^{\frac hQ |{V}|} 
\exp\left(  \sum_{ C_i| \dot \cup_i {\ca C}_i= \ca C } \frac{h^2(Q-1)}{2 Q^2} | C_i|^2 + \ca O(h^3) \right).\qquad 
\eea
where 
\bea
|\ca C| &=&  \mbox{number of edges in } \ca C \\ 
|| \ca C || &=& \mbox{number of connected components in } \ca C \\
|\ca V | &=& \mbox{number of vertices in  the graph}.
\eea

\subsection{Sampling cluster configurations from spin configurations}
\label{s:Sampling cluster configurations from spin configurations}
According to the arguments given above, 
 cluster configurations can be sampled from spin   configurations:
 \begin{enumerate}
\item[(i)] sample a spin configuration with the   weight $\rme^{- {\cal H}^{\rm Potts}_Q[\vec s;\vec h]}$,
\item[(ii)] identify regions of equal spin as a {\em spin domain}. We consider a  bond between two equal spins to belong to this spin domain. 
\item[(ii)]  for each such bond  inside a  spin domain,  remove it with   probability 
$\rme^{-J}$. 
\item[(iv)] each spin domain is by this construction decomposed into one, or several,  clusters (i.e.\ connected components) of the FK expansion. 
\end{enumerate}
An important result of this construction is that the clusters of the FK expansion  live inside the spin domains.

\subsection{An exact lattice action for the cluster expansion}
Let us start from \Eq{35}, 
\be
\ca Z[\vec h] = \sum_{\{ \vec s_x\}}  \sum_{\ca C}  \prod_{( x,y)\in \ca C}   (\rme^J {-}1)\delta_{\vec s_x ,\vec s_y}  \rme^{\sum_x \vec h_x \vec s_x },
\label{35-bis}
\ee
which we rewrite as 
\be
\ca Z[\vec h] = \sum_{\{ \vec s_x\}}  \sum_{\ca C}   \prod_{x} \rme^{ \vec h_x \vec s_x } \prod_{y }   \beta_{xy}\delta_{\vec s_x, \vec s_y}  ,
\label{35-c}
\ee
Here 
\be
\beta_{xy} = \beta_{yx}= \left\{ { (\rme^J {-}1) \mbox{ for } (x,y) \in \ca C \atop 0 \qquad \mbox{ else} \qquad }\right. .
\ee
We claim that \Eq{35-bis} is produced by the path integral over all $Y_x^\alpha$ and $\tilde Y_x^\a$, with   action  (weight)
\be\label{S-magic}
\rme^{-\ca S} =  \prod_{x} \left\{ \rme^{-    \sum_{\alpha}  {{\tilde Y}^{\alpha}_x  Y^{\alpha}_x }  } \bigg[   \sum_{\alpha}\rme^{h^\alpha_x+Y^{\alpha}_x} \bigg] \right\}
\prod_{x,y  } \sqrt{  1+   \sum_{\alpha} \beta_{xy} {\tilde Y}^{\alpha}_x {\tilde Y}^{\alpha}_y } . \qquad 
\ee
\underline{Proof}: 
The  first term in the   action is a term   $\sum_{\alpha}    {{\tilde Y}^{\alpha}_x  Y^{\alpha}_x } $ for each site $x$.
This term is introduced to have a Gaussian measure on each site, with expectation values 
\be\label{Gaussian=theory}
\left<   {{\tilde Y}^{\alpha}_x  Y^{\beta}_y } \right>_0 = \delta^{\alpha\beta} \delta_{xy}, \qquad 
\left<   {{\tilde Y}^{\alpha}_x   } \right>_0 = \left<   {{ Y}^{\alpha}_x    } \right>_0 =\left<   {{\tilde Y}^{\alpha}_x  \tilde Y^{\beta}_y } \right>_0 = \left<   {{ Y}^{\alpha}_x  Y^{\beta}_y } \right>_0= 0. 
\ee
The last term gives a factor of ${  1+   \sum_{\alpha} \beta_{xy} {\tilde Y}^{\alpha}_x {\tilde Y}^{\alpha}_y }$ per edge. The   square root in \Eq{S-magic} corrects for the fact that the product contains both the terms $(x,y)$ and $(y,x)$. Expanding at $x$ in powers of $\beta_{xy}$ 
generates all possible terms in the cluster expansion, and 
each bond in the cluster expansion appears with a factor of $\beta_{xy}$. (Note that the   coupling $J$ on bond $(x,y)$ may differ from bond to bond, and our derivation remains valid for random-bond models).

The next thing to achieve is to contract the $\tilde Y^\alpha_x$ fields, for a given term in the cluster expansion. Due to the rules \eq{Gaussian=theory}, the only available contractions are with the term $\sum_{\alpha}\rme^{h^\alpha_x+Y^{\alpha}_x}$. Writing down   $\tilde Y^\alpha_x$ for each factor of $\beta_{xy}$,  we need to evaluate 
\be
\sum_\alpha \left< \rme^{h^\alpha_x+Y^\alpha_x} \tilde Y^{\beta}_x \tilde Y^{\gamma}_x \ldots \tilde Y^{\delta}_x \right>
= \sum _\alpha \rme^{h^\alpha_x} \delta^{\alpha \beta}\delta^{\alpha \gamma} \ldots \delta^{\alpha \delta}.
\ee
Thus whenever the cluster-expansion contains the term $\beta_{xy}$, our action forces the $Y^\alpha$ and $\tilde Y^\alpha$ fields to have the same index $\alpha$. This gives a factor of $\sum_\alpha\rme^{\sum_{x\in \ca C} h^\alpha_x }$  per cluster $\ca C$. This completes the proof.
For the most interesting case of vanishing magnetic field,  $h=0$, this reduces to  $\sum_\alpha = Q$.

\subsection{Expansion of the cluster action in the fields}
We can expand \Eq{S-magic} in powers of the field. This will be relevant to access the critical theory in $d=6-\epsilon$ dimensions. With this in mind, we expand the action up to third order, putting $h_x^\alpha \to 0$. We start with the 
auxiliary formula
\be
\sum_{\alpha}\rme^{Y^{\alpha}_x} = Q + \sum_{\alpha}  Y^{\alpha}_x + \frac12  \sum_{\alpha}  (Y^{\alpha}_x)^2 + \frac1{3!}  \sum_{\alpha}  (Y^{\alpha}_x)^3 + ...
\ee
This implies that the corresponding contribution to the action $\ca S$ from \Eq{S-magic} reads 
\bea
&& -\ln \left( \sum_{\alpha}\rme^{Y^{\alpha}_x}\right) = -\ln Q - \frac1Q \sum_{\alpha}  Y^{\alpha}_x \nn\\
&&\qquad + \frac1{2} \Big[\frac1Q \sum_{\alpha}  Y^{\alpha}_x \Big]^2  -    \frac1{2Q} \sum_{\alpha}  (Y^{\alpha}_x)^2\nn\\
&&\qquad - \frac1{3} \Big[\frac1Q \sum_{\alpha}  Y^{\alpha}_x \Big]^3+ \frac1{2} \Big[\frac1Q \sum_{\alpha}  (Y^{\alpha}_x)^2 \Big] \Big[\frac1Q \sum_{\alpha}  Y^{\alpha}_x \Big] -    \frac1{6Q} \sum_{\alpha}  (Y^{\alpha}_x)^3  + \ca O(Y^4).
\eea
The next contribution to $\ca S$ from \Eq{S-magic} reads
\bea
&&-\ln \prod_{y  } \sqrt{  1+   \sum_{\alpha} \beta_{xy} {\tilde Y}^{\alpha}_x {\tilde Y}^{\alpha}_y } = - \half \sum_{y}    \sum_{\alpha} \beta_{xy} {\tilde Y}^{\alpha}_x {\tilde Y}^{\alpha}_y +...\nn\\
&& = \frac14  \sum_{y}    \sum_{\alpha} \beta_{xy} \bigg[ \Big({\tilde Y}^{\alpha}_x-{\tilde Y}^{\alpha}_y\Big)^2 - ({\tilde Y}^{\alpha}_x)^2-({\tilde Y}^{\alpha}_y)^2 \bigg] + ...
\nn\\
&& = \frac14     \sum_{\alpha} \sum_{y} \beta_{xy}  \Big({\tilde Y}^{\alpha}_x-{\tilde Y}^{\alpha}_y\Big)^2  -\frac12  \Big(\sum_{y}  \beta_{xy}  \Big)  \sum_{\alpha}  ({\tilde Y}^{\alpha}_x)^2 + ..., 
\eea
where we used that $\beta_{xy}=\beta_{yx}$ and this contribution is summed over $x$. 
Therefore, up to a constant, 
\bea\label{SYYtilde}
{\ca S}[\tilde Y,Y] &=&  \sum_x \bigg\{ \sum_{\alpha}  {\tilde Y}^{\alpha}_x  Y^{\alpha}_x- \frac1Q \sum_{\alpha}  Y^{\alpha}_x \nn\\
&&\qquad + \frac1{2} \Big[\frac1Q \sum_{\alpha}  Y^{\alpha}_x \Big]^2  -    \frac1{2Q} \sum_{\alpha}  (Y^{\alpha}_x)^2\nn\\
&&\qquad - \frac1{3} \Big[\frac1Q \sum_{\alpha}  Y^{\alpha}_x \Big]^3+ \frac1{2} \Big[\frac1Q \sum_{\alpha}  (Y^{\alpha}_x)^2 \Big] \Big[\frac1Q \sum_{\alpha}  Y^{\alpha}_x \Big] -    \frac1{6Q} \sum_{\alpha}  (Y^{\alpha}_x)^3 + \ca O(Y^4)\nn\\
&& \qquad +\frac14     \sum_{\alpha} \sum_{y} \beta_{xy}  \Big[{\tilde Y}^{\alpha}_x-{\tilde Y}^{\alpha}_y\Big]^2  -\frac12  \Big[\sum_{y}  \beta_{xy}  \Big]  \sum_{\alpha}  ({\tilde Y}^{\alpha}_x)^2   + \ca O (\tilde Y)^4\bigg\}
\eea
\subsection{Integrating out ${\tilde Y}^{\alpha}$}
Since ${\tilde Y}^{\alpha}_x$ only appears quadratically   in the action \eq{SYYtilde}, we can   integrate it out.
(Higher-order terms   have to be dealt with when including terms of order $\tilde Y^4$.)
To do so, we take the saddle point 
\be
0 = \frac{\rmd {\ca S}[\tilde Y,Y] }{\rmd {\tilde Y}^{\alpha}_x} = Y^{\alpha}_x -      \sum_{y} \beta_{xy}  
\Big[{\tilde Y}^{\alpha}_y-{\tilde Y}^{\alpha} _x\Big] 
-  \Big[\sum_{y}  \beta_{xy}  \Big]    {\tilde Y}^{\alpha}_x  ,
\ee
where   again we used $\beta_{xy}=\beta_{yx}$.
This can be rewritten as 
\be\label{78}
 Y^{\alpha}_x = \nabla^2_\beta {\tilde Y}^{\alpha}_x +M  {\tilde Y}^{\alpha}_x
\ee
where
\be
\nabla^2_\beta {\tilde Y}^{\alpha}_x :=  \sum_{y} \beta_{xy}  \Big[{\tilde Y}^{\alpha}_y-{\tilde Y}^{\alpha}_x\Big] 
\ee
is the lattice Laplacian.    There is  a    mass term $m^2=M$, 
\be
M  :=  \sum_{y}  \beta_{xy} = (\rme^J -1) \mbox{coord},
\ee
where $\mbox{coord}$ is the coordination number, i.e.\ the number of nearest neighbors. 
(The signs are for $J>0$, i.e.\ ferromagnetic couplings.)
Then \Eq{78} can be inverted, 
\be
{\tilde Y}^{\alpha}_x =  \frac{1}{M + \nabla^2_\beta }Y^{\alpha}_x\approx \left[   \frac{1}{M} - \frac{\nabla^2_\beta } {M^2}+ ...\right] Y^{\alpha}_x.
\ee
This  gives 
\bea
\!\!\!{\ca S}[Y] &=&  \sum_x \bigg\{ \sum_{\alpha}  \frac1{2M} (Y^{\alpha}_x)^2    + \frac{1}{4M^2 }      \sum_{\alpha} \sum_{y} \beta_{xy}  \Big[Y^{\alpha}_x-Y^{\alpha}_y\Big]^2  - \frac1Q \sum_{\alpha}  Y^{\alpha}_x \nn\\
&&\qquad + \frac1{2} \Big[\frac1Q \sum_{\alpha}  Y^{\alpha}_x \Big]^2  -    \frac1{2Q} \sum_{\alpha}  (Y^{\alpha}_x)^2\nn\\
&&\qquad - \frac1{3} \Big[\frac1Q \sum_{\alpha}  Y^{\alpha}_x \Big]^3+ \frac1{2} \Big[\frac1Q \sum_{\alpha}  (Y^{\alpha}_x )^2\Big] \Big[\frac1Q \sum_{\alpha}  Y^{\alpha}_x \Big] -    \frac1{6Q} \sum_{\alpha}  (Y^{\alpha}_x)^3 + \ca O(Y^4)\bigg\}. \qquad
\eea
Note that the contribution from the determinant in the integration over $\tilde Y$ is independent of $Y$, thus a constant which can be neglected.

\subsection{Decomposition   into scalar and traceless parts}

The next step is to decompose $Y_{\alpha}$ into irreducible representations of the symmetric group (which exchanges the flavours of the Potts spin), see e.g.\ \cite{Cardy1999,VasseurJacobsenSaleur2012,NahumPhD}, 
\be\label{Y2rep}
Y^{\alpha}_x = \Phi^{\alpha}_x + S_x ,\qquad \sum_{\alpha} \Phi^{\alpha}_x =0.
\ee
$\Phi^\alpha$ is the vector representation with Young tableau $[Q-1,1]$ and dimension $Q-1$,
while $S$ is the scalar representation with Young tableau $[Q]$ and dimension 1. Both are irreducible for $Q > 1$.
This gives for the action
\bea\label{47}
{\ca S}[S,\Phi] &=&  \sum_x \bigg\{ \sum_{\alpha}  \frac1{2M} (\Phi^{\alpha}_x)^2    +  \frac{1}{4M^2 }      \sum_{\alpha} \sum_{y} \beta_{xy}  \big(\Phi^{\alpha}_x-\Phi^{\alpha} _y\big)^2 \nn\\
&&\qquad  +   \frac Q{2M } S_x^2    +  \frac{Q}{4M^2  }       \sum_{y} \beta_{xy}  \big(S_x-S_y\big)^2  - S_x \nn\\
&&\qquad  -    \frac1{2Q} \sum_{\alpha}  (\Phi^{\alpha}_x)^2     -    \frac1{6Q} \sum_{\alpha}   (\Phi^{\alpha}_x)^3 + \ca O(\Phi_{\alpha}^4)\bigg\}.
\eea
Note that the terms non-linear  in $S_x$ which come from $\sum_\alpha \rme^{Y_x^\a}$ have all canceled. This property is   {\em exact and holds to all orders}. It can be traced back to 
\be\label{trick}
\ln \Big(\sum_{\alpha} \rme^{Y^{\alpha}_x} \Big) = \ln \Big(\sum_{\alpha} \rme^{\Phi^{\alpha}_x+S_x} \Big) \equiv  \ln \Big(\rme^{S_x} \sum_{\alpha} \rme^{\Phi^{\alpha}_x} \Big)  \equiv S_x+ \ln \Big(\sum_{\alpha} \rme^{\Phi^{\alpha}_x} \Big) . 
\ee
Thus there are no non-linear terms in $S_x$ from the vertex! The only   terms of order larger than two that may  appear are from the quartic term $ \sum_{xy}\sum_\alpha (\beta_{xy} \tilde Y^{\alpha}_x \tilde Y^{\alpha}_y)^2$ (or higher).

Let us write the action with these simplifications. 
The mode $S_x$ is   massive with squared mass $Q/M$ (second line of \Eq{47}), and   expectation 
\be\label{87}
\left< S_x \right> =  \frac {M}Q.
\ee
However $S_x$ decouples in\Eq{47}.   The remaining action for $\Phi^{\alpha}_x$ reads
\be
{\ca S}[\Phi] =  \sum_x \bigg\{ \sum_{\alpha}  \left( \frac1{M}-    \frac1{Q} \right) \frac1{2}(\Phi^{\alpha}_x)^2    +{  \frac{1}{4M^2 }     } \sum_{\alpha} \sum_{y} \beta_{xy}  \big(\Phi^{\alpha}_x-\Phi^{\alpha}_y\big)^2    -    \frac1{6Q} \sum_{\alpha}   (\Phi^{\alpha}_x)^3 + \ca O(\Phi_{\alpha}^4)\bigg\}.
\ee
This is a cubic field theory, for any value of $Q\neq 0$. 

The  modes $\Phi_{\alpha}$ have a bare mass
\be
m_0^2 = \frac1{
M} -\frac1 Q  \quad \Longrightarrow \quad\Phi^{\alpha} \mbox{   is massless in the bare theory for } M \equiv \sum_y \beta_{xy}\equiv (\rme^J{-}1)\mbox{coord}= Q.
\ee

\subsection{Local observables}
A key information  to know is which cluster site $x$ belongs to.  This can be achieved by dropping the sum at site $x$. 
Formally, the operator which tells whether site $x$ is in a cluster of color $\beta$ is 
\be
 \ca O^\beta_x := \frac{\rme^{Y^{\beta}_x}}{ \sum_{\alpha}\rme^{Y^{\alpha}_x} } .
\ee
The  denominator   takes out the corresponding term from the action \eq{S-magic}, and replaces it by the same term without the sum.
As constructed, 
\be \sum_\b \ca O^{\beta}_x = 1.
\ee
To access the  probability that different sites are in the same cluster, 
we define its connected part, 
\be
 \hat {\ca O}^\beta_x :=  \ca O^\beta_x - \left<  \ca O^\beta_x \right> =  \ca O^\beta_x - \frac 1 Q. 
\ee
Expressed in terms of $\Phi^\alpha$ and $S$, this reads
\be
 \hat {\ca O}^\beta_x  = \frac{\rme^{S+\Phi^{\beta}_x}}{ \sum_{\alpha}\rme^{S+\Phi^{\alpha}_x} }-\frac1Q=
 \frac{\rme^{\Phi^{\beta}_x}}{ \sum_{\alpha}\rme^{\Phi^{\alpha}_x} }-\frac1Q = \frac{\Phi^\b_x}Q + \frac{1}{2Q}\left[ {(\Phi^\b_x)^2} - \frac1Q \sum_\alpha {(\Phi^\a_x)^2} \right]  + \ca O(\Phi^\a_x)^3
\ee
The  probability that $n$ sites $x_1\ldots x_n$ are in the same cluster is 
proportional to 
\be
\left<  \hat {\ca O}^\beta_{x_1}   \hat {\ca O}^\beta_{x_2} \ldots    \hat {\ca O}^\beta_{x_n}\right> =  \frac1{Q^n}
\left< \Phi^\beta_{x_1}  \Phi^\beta_{x_2} \ldots    \Phi^\beta_{x_n}\right> +\mbox{higher-order terms}.
\ee

\section{Renormalization: $6-\epsilon$ expansion}
\label{s:Renormalization: 6-epsilon expansion}
The renormalization group for the $Q$-state Potts model is usually performed in momentum space.
Here we present this standard calculation in position space.  The advantage is that the 3-point function encoding the structure function is then easily evaluated. 

\subsection{Relevant relations and normalizations}
Using the definition of the $\Gamma$-function, one first shows that 
\be
\frac{1}{|x|^{2a}}= \int_{0}^\infty  \frac{s^{a-1} e^{-s x^2}}{\Gamma (a)} \,\rmd s.
\ee
This allows us to Fourier transform power laws according to 
\be\label{110}
\int \rmd^d x\, \frac{\rme^{i \vec k \vec x}}{|x|^{2a}} = \pi^{d/2} \left| \frac k2\right| ^{2a-d}\frac{ \Gamma
   \left(\frac{d}{2}-a\right)}   {\Gamma (a)} .
\ee
\Eq{110} for $a=(d-2)/2$ implies  
\be
\int \rmd^d x\, \frac{\rme^{i \vec k \vec x}}{|x|^{d-2}} = \frac1{ \left| k\right| ^{2}}\frac{ 4 \pi^{d/2} }   {\Gamma (\frac{d-2}2)} 
=\frac{S_d(d-2)}{k^2}.
\ee
In order to reduce as much as possible geometric factors,   we introduced the $d$-dimensional volume element
\be
S_d :=  \frac{2\pi^{d/2}}{\Gamma(d/2)}.
\ee
A useful relation is 
\be\label{109}
\frac1{S_d}\int \rmd^d y\, \frac1{|y|^{2a}|x-y|^{2b}} =\frac{\Gamma(\frac d2) \Gamma
   \left(\frac{d}{2}-a\right)
   \Gamma
   \left(\frac{d}{2}-b\right) \Gamma
   \left(a+b-\frac{d}{2}\right
   )}{2\Gamma (a) \Gamma (b)
   \Gamma (d-a-b)} \,|x|^{d-2 (a+b)}.
\ee
This is proven by going to momentum space w.r.t.\ to $y$ and $x-y$,   multiplying the two momentum dependent 
functions, and transforming back. 

Following CFT conventions, 
field theory is constructed with propagators normalized such that 
\be
\left< \Phi_\alpha(x) \Phi_\beta(y) \right> = |x-y|^{2-d} \left( \delta _{\alpha\beta}-\frac1 Q\right).
\ee
The {\em bare} Lagrangian in these normalizations is 
\be\label{L}
\ca L  = \frac1{(d-2)S_d} \sum_\alpha\frac12 \left[\nabla \Phi_\alpha^0(x)\right]^2+ \frac1{S_d} \frac{g_0}{3!} \sum_\alpha \Phi_\alpha^0(x)^3.
\ee
The index $\alpha$ is the field index, while the index $0$ refers to bare quantities.

\subsection{Renormalization scheme}

Following  the standard field-theoretic scheme \cite{ZiaWallace1975,Zinn-Justin2021,Amit1,Amit1976}, 
renormalization is performed by introducing   RG-factors, 
\be\label{action-64}
\ca L  =  \frac{Z }{(d-2)S_d} \sum_\alpha\frac12 \left[\nabla \Phi_\alpha(x)\right]^2+ \frac1{S_d} \frac{g Z_g  L^{-\frac\epsilon2}}{3!}   \sum_\alpha \Phi_\alpha(x)^3.
\ee
Here all quantities are renormalized, contrary to \Eq{L}, where they are bare. 
The relation between bare and renormalized quantities is  
\bea
g_0 &=& g Z_g  Z^{- \frac32} L^{-\frac\epsilon 2} ,\\
\Phi_\a^{0} &=& \sqrt{Z} \Phi_\alpha  .
\eea
This yields the $\beta$ function $\beta_g(g)$, full field dimension $\Delta$, and anomalous exponent $\eta$ as
\bea
\label{beta-def}
\beta_g(g) &=& L \partial_L g = \frac{\epsilon}2 \frac{ g }{1+ g\partial_g \ln (Z_g Z^{-3/2})} ,\\
\label{Delta-def}
\Delta &=& \frac{d-2}2 + \gamma_\phi  , \\
\label{eta-def}
\eta &=& 2 \gamma_\phi =  - L \partial _L \ln (Z) = - \beta_g(g)\partial_g \ln (Z).
\eea

\subsection{Vertex renormalization}

There are two types of  correction at 1-loop-order. The first is a correction to $\Gamma^{(3)}$,  the vertex.  
Graphically, it can be written as 
\bea\label{60}
\fboxsep0mm
\parbox{1.3cm}{{\begin{tikzpicture}[scale=1]
\coordinate (x1) at (0,0) ;
\coordinate (x2) at  (0.7,0) ;
\coordinate (x3) at  (0.35,0.5) ;
\coordinate (x1p) at (-.3,-0.1) ;
\coordinate (x2p) at  (1,-.1) ;
\coordinate (x3p) at  (0.35,0.8) ;
\node [below] at (x1)  {$x$} ;
\node [below] at (x2)  {$y$} ;
\node [above,left] at (x3)  {$z$} ;
\fill (x1) circle (1.5pt);
\fill (x2) circle (1.5pt);
\fill (x3) circle (1.5pt);
\draw [blue] (x1) -- (x2) -- (x3) -- (x1);
\draw [blue] (x1) -- (x1p) ;
\draw [blue] (x2) -- (x2p) ;
\draw [blue] (x3) -- (x3p) ;
\end{tikzpicture}}}
&=& \parbox{1.3cm}{{\begin{tikzpicture}[scale=1]
\coordinate (x1) at (0,0) ;
\coordinate (x2) at  (0.7,0) ;
\coordinate (x3) at  (0.35,0.5) ;
\coordinate (x1p) at (-.2,-0.1) ;
\coordinate (x2p) at  (0.9,-.1) ;
\coordinate (x3p) at  (0.35,0.75) ;
\fill (x1) circle (1.5pt);
\fill (x2) circle (1.5pt);
\fill (x3) circle (1.5pt);
\draw [blue,thick] (x1) -- (x2) -- (x3) -- (x1);
\draw [blue] (x1) -- (x1p) ;
\draw [blue] (x2) -- (x2p) ;
\draw [blue] (x3) -- (x3p) ;
\end{tikzpicture}}}
-\frac1Q \Bigg[ ~\parbox{1.3cm}{{\begin{tikzpicture}[scale=1]
\coordinate (x1) at (0,0) ;
\coordinate (x2) at  (0.7,0) ;
\coordinate (x3) at  (0.35,0.5) ;
\coordinate (x1p) at (-.2,-0.1) ;
\coordinate (x2p) at  (0.9,-.1) ;
\coordinate (x3p) at  (0.35,0.75) ;
\fill (x1) circle (1.5pt);
\fill (x2) circle (1.5pt);
\fill (x3) circle (1.5pt);
\draw [blue,thick] (x1) -- (x2) -- (x3) ;
\draw [blue,thick,dashed] (x1) -- (x3) ;
\draw [blue] (x1) -- (x1p) ;
\draw [blue] (x2) -- (x2p) ;
\draw [blue] (x3) -- (x3p) ;
\end{tikzpicture}}}+\parbox{1.3cm}{{\begin{tikzpicture}[scale=1]
\coordinate (x1) at (0,0) ;
\coordinate (x2) at  (0.7,0) ;
\coordinate (x3) at  (0.35,0.5) ;
\coordinate (x1p) at (-.2,-0.1) ;
\coordinate (x2p) at  (0.9,-.1) ;
\coordinate (x3p) at  (0.35,0.75) ;
\fill (x1) circle (1.5pt);
\fill (x2) circle (1.5pt);
\fill (x3) circle (1.5pt);
\draw [blue,thick]  (x2) -- (x3) -- (x1);
\draw [blue,thick,dashed] (x1) -- (x2) ;
\draw [blue] (x1) -- (x1p) ;
\draw [blue] (x2) -- (x2p) ;
\draw [blue] (x3) -- (x3p) ;
\end{tikzpicture}}} 
+\parbox{1.3cm}{{{\begin{tikzpicture}[scale=1]
\coordinate (x1) at (0,0) ;
\coordinate (x2) at  (0.7,0) ;
\coordinate (x3) at  (0.35,0.5) ;
\coordinate (x1p) at (-.2,-0.1) ;
\coordinate (x2p) at  (0.9,-.1) ;
\coordinate (x3p) at  (0.35,0.75) ;
\fill (x1) circle (1.5pt);
\fill (x2) circle (1.5pt);
\fill (x3) circle (1.5pt);
\draw [blue,thick] (x3) -- (x1) -- (x2) ;
\draw [blue,thick,dashed] (x3) -- (x2) ;
\draw [blue] (x1) -- (x1p) ;
\draw [blue] (x2) -- (x2p) ;
\draw [blue] (x3) -- (x3p) ;
\end{tikzpicture}}}}
\Bigg] \nn\\
&& + \frac1{Q^{2}}\Bigg[ ~\parbox{1.3cm}{{{\begin{tikzpicture}[scale=1]
\coordinate (x1) at (0,0) ;
\coordinate (x2) at  (0.7,0) ;
\coordinate (x3) at  (0.35,0.5) ;
\coordinate (x1p) at (-.2,-0.1) ;
\coordinate (x2p) at  (0.9,-.1) ;
\coordinate (x3p) at  (0.35,0.75) ;
\fill (x1) circle (1.5pt);
\fill (x2) circle (1.5pt);
\fill (x3) circle (1.5pt);
\draw [blue,thick] (x1) -- (x2);
\draw [blue,thick,dashed] (x1)--(x3) -- (x2);
\draw [blue] (x1) -- (x1p) ;
\draw [blue] (x2) -- (x2p) ;
\draw [blue] (x3) -- (x3p) ;
\end{tikzpicture}}}}+\parbox{1.3cm}{{{\begin{tikzpicture}[scale=1]
\coordinate (x1) at (0,0) ;
\coordinate (x2) at  (0.7,0) ;
\coordinate (x3) at  (0.35,0.5) ;
\coordinate (x1p) at (-.2,-0.1) ;
\coordinate (x2p) at  (0.9,-.1) ;
\coordinate (x3p) at  (0.35,0.75) ;
\fill (x1) circle (1.5pt);
\fill (x2) circle (1.5pt);
\fill (x3) circle (1.5pt);
\draw [blue,thick] (x3) -- (x2);
\draw [blue,thick,dashed] (x2)--(x1)--(x3);
\draw [blue] (x1) -- (x1p) ;
\draw [blue] (x2) -- (x2p) ;
\draw [blue] (x3) -- (x3p) ;
\end{tikzpicture}}}} +\parbox{1.3cm}{{{\begin{tikzpicture}[scale=1]
\coordinate (x1) at (0,0) ;
\coordinate (x2) at  (0.7,0) ;
\coordinate (x3) at  (0.35,0.5) ;
\coordinate (x1p) at (-.2,-0.1) ;
\coordinate (x2p) at  (0.9,-.1) ;
\coordinate (x3p) at  (0.35,0.75) ;
\fill (x1) circle (1.5pt);
\fill (x2) circle (1.5pt);
\fill (x3) circle (1.5pt);
\draw [blue,thick] (x3) -- (x1);
\draw [blue,thick,dashed] (x1)--(x2) -- (x3);
\draw [blue] (x1) -- (x1p) ;
\draw [blue] (x2) -- (x2p) ;
\draw [blue] (x3) -- (x3p) ;
\end{tikzpicture}}}}
\Bigg] 
- \frac1{Q^3}~\parbox{1.3cm}{{{\begin{tikzpicture}[scale=1]
\coordinate (x1) at (0,0) ;
\coordinate (x2) at  (0.7,0) ;
\coordinate (x3) at  (0.35,0.5) ;
\coordinate (x1p) at (-.2,-0.1) ;
\coordinate (x2p) at  (0.9,-.1) ;
\coordinate (x3p) at  (0.35,0.75) ;
\fill (x1) circle (1.5pt);
\fill (x2) circle (1.5pt);
\fill (x3) circle (1.5pt);
\draw [blue,thick,dashed] (x1) -- (x2) -- (x3) -- (x1);
\draw [blue] (x1) -- (x1p) ;
\draw [blue] (x2) -- (x2p) ;
\draw [blue] (x3) -- (x3p) ;
\end{tikzpicture}}}}  \nn\\
&=& \Big(1-\frac3Q\Big) \sum_\alpha \Phi_\alpha^3 .
\eea
A thick solid line signifies the term $\delta^{\alpha\beta}$ in the propagator, whereas the dashed line does not force the indices to be equal; the accompanging factor of $1/Q$ is written explicitly. 
The   diagrams in the second  line vanish due to $\sum_\alpha \Phi_\alpha=0$, and are not   reported in the final result.  Including all combinatorial factors, the perturbative result for the cubic vertex $\Gamma^{(3) }$ is 
\be
 \Gamma^{(3) } = g_0  \left[ 1 + \frac1{3!} \left(\frac{g_0}{3!}\right)^2 \times 3^3 \times 4 \times 2 \times \left(1-\frac 3 Q\right) \ca I L^\epsilon \right]
= g_0+   g_0^3 \ca I L^{\epsilon}  \left(1-\frac 3 Q\right).
\ee
The combinatorial factors are: $1/3!$ from the third-order expansion of the exponential function; a factor of $g_0/3!$ for each additional vertex as written in the action \eq{action-64}; a factor of $3$ per vertex for choosing the uncontracted leg; a factor of $4$ for contracting a first chosen leg; a factor of $2$ for the remaining contraction.
The diagram having 3 propagators, one can  for each of them choose the contribution proportional to the $\delta$-function for the indices, or in any of the three  use the term without a $\delta$-function, for a total of $(1-3/Q)$; 
using for two or three propagators the term without the $\delta$-function would break the connectedness of the diagram. 
The final factor is  the integral. 
Regularized at scale $L$ it is  
\bea
\ca I L^\epsilon &=& \frac1{S_d^2}\int_x  \int_y\Theta(|x|<L) \Big( |y||y-x| |x| \Big)^{-2\Delta} \nn\\
&=& \frac{\Gamma(\frac d2) \Gamma
   \left(\frac{d}{2}-\Delta
   \right)^2 \Gamma \left(2 \Delta
   -\frac{d}{2}\right)
   }{2\Gamma (\Delta )^2 \Gamma
   (d-2 \Delta )}
   \frac1{S_d}\int_x \Theta(|x|<L)  {|x|}^{{d} -6 \Delta
   }.
\eea
We used \Eq{109}. Note that we could equivalently   put a cutoff on both $x$ and $y$, see \cite{WieseDavid1995,WieseDavid1997,WieseHabil}.
The remaining integral gives
\be
\frac1{S_d}\int_x \Theta(|x|<L)  {|x|}^{{d} -6 \Delta} =  \int_0^L \frac{\rmd x}{x} x^{2d-6\Delta}=   \int_0^L \frac{\rmd x}{x} x^{6-d} =   \frac{L^\epsilon}{\epsilon}.
\ee
Therefore with $\Gamma(d/2)\simeq 2 $,  and   the remaining $\Gamma$'s evaluating to 1, 
\be
\ca I =     \frac 1 \epsilon + \ca O(\epsilon^0)  .
\ee
This identifies 
\be
Z_g = 1 - \frac {g^2} \epsilon \left(1-\frac 3Q\right)+ \ca O(g^4).
\ee

\subsection{Renormalization  of the elastic term}
The second contribution is the renormalization of $\Gamma^{(2)}$,  the elastic term (wave-function renormalization).
Similarly to what has been done in \Eq{60}, the 
group-theoretical factor is 
\be
{ {\parbox{1.cm}{{\begin{tikzpicture}[scale=1]
\coordinate (x1) at (0,0) ;
\coordinate (x2) at  (0.6,0) ;
\coordinate (x1p) at  (-.2,0) ;
\coordinate (x2p) at  (0.8,0) ;
\fill (x1) circle (1.5pt);
\fill (x2) circle (1.5pt);
\draw [blue] (.3,0) circle (3mm);
\draw [black] (x1) -- (x1p);
\draw [black] (x2) -- (x2p);
\end{tikzpicture}}}} 
=
{\parbox{1.cm}{{\begin{tikzpicture}[scale=1]
\coordinate (x1) at (0,0) ;
\coordinate (x2) at  (0.6,0) ;
\coordinate (x1p) at  (-.2,0) ;
\coordinate (x2p) at  (0.8,0) ;
\fill (x1) circle (1.5pt);
\fill (x2) circle (1.5pt);
\draw [blue,thick] (.3,0) circle (3mm);
\draw [black] (x1) -- (x1p);
\draw [black] (x2) -- (x2p);
\end{tikzpicture}}}}
-\frac 2 Q\,
{\parbox{1.cm}{{\begin{tikzpicture}[scale=1]
\coordinate (x1) at (0,0) ;
\coordinate (x2) at  (0.6,0) ;
\coordinate (x1p) at  (-.2,0) ;
\coordinate (x2p) at  (0.8,0) ;
\fill (x1) circle (1.5pt);
\fill (x2) circle (1.5pt);
\draw [thick,blue,dashed] (.6,0) arc (0:180:3mm);
\draw [thick,blue] (0,0) arc (-180:0:3mm);
\draw [black] (x1) -- (x1p);
\draw [black] (x2) -- (x2p);
\end{tikzpicture}}}}}
+\frac1{Q^2}\,
{\parbox{1.cm}{{\begin{tikzpicture}[scale=1]
\coordinate (x1) at (0,0) ;
\coordinate (x2) at  (0.6,0) ;
\coordinate (x1p) at  (-.2,0) ;
\coordinate (x2p) at  (0.8,0) ;
\fill (x1) circle (1.5pt);
\fill (x2) circle (1.5pt);
\draw [blue,thick,dashed] (.3,0) circle (3mm);
\draw [black] (x1) -- (x1p);
\draw [black] (x2) -- (x2p);
\end{tikzpicture}}}}.
\ee
Due to $\sum_\alpha \Phi_\alpha=0$, 
the last term does not contribute. Therefore, this relation can be simplified to 
\be
{ {\parbox{1.cm}{{\begin{tikzpicture}[scale=1]
\coordinate (x1) at (0,0) ;
\coordinate (x2) at  (0.6,0) ;
\coordinate (x1p) at  (-.2,0) ;
\coordinate (x2p) at  (0.8,0) ;
\fill (x1) circle (1.5pt);
\fill (x2) circle (1.5pt);
\draw [blue] (.3,0) circle (3mm);
\draw [black] (x1) -- (x1p);
\draw [black] (x2) -- (x2p);
\end{tikzpicture}}}} } =  {\Big(1-\frac2Q\Big)}  \times   \propdiag~.
\ee
Writing explicitly the fields to keep track of the derivatives,  the contribution to $\Gamma^{(2)}$ reads
\bea
\delta \Gamma^{(2)}  &=& - \left(\frac{g_0}{3! S_d}\right)^2\times \frac1{2!}  \times 3^2 \times 2 \times \left(1-\frac2Q \right) \int_y \sum_\alpha\Phi_\alpha(x) \Phi_\alpha(y) \frac{1}{|x-y|^{2(d-2)}} \nn\\
&=& -  \frac{g_0^2 }{  S_d}   \frac14   \left(1-\frac2Q \right) \frac1{S_d}\int_y \sum_\alpha \Phi_\alpha(x) \Big\{  \Phi_\alpha(x) +(y-x)\nabla \Phi_\alpha(x) + \frac12 
  \left[ (y-x)\nabla \right]^2 \Phi_\alpha(x) + ... \Big\}\nn\\
  && \qquad \qquad \qquad \qquad \qquad   \times \frac{1}{|x-y|^{2(d-2)}}  \nn\\
&=& -  \frac{g_0^2 }{  S_d}   \frac14   \left(1-\frac2Q \right) \frac1{S_d}\int_y \Phi_\alpha(x) \Big\{  \Phi_\alpha(x) + \frac1{2d} 
   (y-x)^2 \nabla^2 \Phi_\alpha(x) + ... \Big\} \frac{1}{|x-y|^{2(d-2)}} . 
\eea
The first $(-1)$ is due to the fact that we have $\rme^{-\int \ca L}$, the $1/2!$ is from the second order in $g_0$, then combinatorial factors, group factors, and the integral. In the third line we used rotational invariance to discard the  first-order term in $\nabla$, and simplify the second term. 
Regrouping and dropping the UV-relevant term gives  
\bea
\delta \Gamma^{(2)}    &=&  \frac{g_0^2 }{  S_d} \times \frac1{4d} \times \left(1-\frac2Q \right) \sum_{\alpha}\frac12 \left[\nabla \Phi_\alpha(x)\right]^2 \frac1{S_d}\int_y 
\frac{|x-y|^2}{|x-y|^{2(d-2)}} \nn\\
&=&  {g_0^2 }   \frac{d-2}{4d}   \left(1-\frac2Q \right) \frac{L^{\epsilon}}{\epsilon} \times \frac1{2(d-2)S_d}  \sum_{\alpha}\left[\nabla \Phi_\alpha(x)\right]^2 + \ca O(\epsilon^0) .
\eea
The last term is the  free Lagrangian density, yielding the field renormalization factor (for $d\to 6$)
\be
Z = 1 -   \frac{g^2}6   \left(1-\frac2Q \right) .
\ee

\subsection{RG functions}
The $\beta$-function as defined in \Eq{beta-def} is
\bea
\beta_g(g) =
\frac{\epsilon}2 \frac{ g }{1+ g\partial_g \ln (Z_g Z^{-3/2})}
&=& \frac{\epsilon}2 g +      \frac{g ^3}2     \left[\frac32 - \frac 5 Q \right]+ \ca O(g^5).
\eea
The fixed point is at 
\be\label{FP-space}
g_*^2 = \frac{-\epsilon }{\frac32 - \frac 5 Q }.
\ee
Note that for $Q< 10/3$ the fixed point is real, while for larger $Q$   it is imaginary. The latter situation contains the Lee-Yang class, to which our RG equations reduce in the limit of $Q\to \infty$. Intuitively this can be understood by remarking that for  $Q\to \infty$ the constraint $\sum_\alpha \Phi^\alpha =0$ becomes negligible, and the Lagrangian \eq{L} reduces to that of $Q$ decoupled Lee-Yang field theories.

The renormalization-group $\eta$ function defined in \Eq{eta-def} reads  
\be\label{eta(g)}
\eta(g) = \frac{g^2}6   \left(1-\frac2Q \right) + \ca O(g^4).
\ee
Evaluated at the fixed point \eq{FP-space},  the exponent $\eta$ becomes  
\be\label{eta-cluster}
\eta(g^{*}) \equiv 2\gamma_\phi=\frac{ \epsilon (Q-2)}{3  (10-3 Q)} + \ca O(\epsilon^{2})\ .
\ee
This agrees with standard treatments \cite{Amit1976,AlcantaraBonfirmKirkhamMcKane1981}.

\subsection{3-point function and structure factor $ C$}
\label{3-point function and structure factor C}
We now consider the 
 3-point function which forces all external indices to be in the same cluster, 
\bea \frac{C}{\left(|x_1-x_2||x_1-x_3||x_2-x_3|\right)^\Delta}
&=&\frac1{S_d}\int_y   \frac{g_*}{\left( |x_1-y| |x_2-y| |x_3-y|\right)^{2\Delta}}\nn\\
&\simeq& g_* \frac{\Gamma(\frac d2)}{2}   \left[\frac{\Gamma(\frac{\Delta}2)}{\Gamma(\Delta)} \right]^3\bigg|_{|x_i-x_j|=1 \forall i\neq j},
\eea
where we used the star-triangle identity. The latter reads
\begin{align}
\label{eq:StarTriangle}
\frac1{S_d}\int \frac{\rmd^dx_4}{
x_{14}^{2\Delta_1}x_{24}^{2\Delta_2}
x_{34}^{2\Delta_3}}=
\frac{\kappa(\Delta_1,\Delta_2,\Delta_3)}{
x_{12}^{\Delta_{12,3}}
x_{13}^{\Delta_{13,2}}
x_{23}^{\Delta_{23,1}}}\ ,
\end{align}
\be
\kappa(\Delta_1,\Delta_2,\Delta_3)=\frac{\Gamma(\frac d2)
\Gamma\left(\frac{\Delta_1+\Delta_2-\Delta_3}{2}\right)
\Gamma\left(\frac{\Delta_1+\Delta_3-\Delta_2}{2}\right)
\Gamma\left(\frac{\Delta_2+\Delta_3-\Delta_1}{2}\right)
}{
2\Gamma(\Delta_1)\Gamma(\Delta_2)\Gamma(\Delta_3) }\simeq 1  \ .
\ee
It holds whenever  $\sum_i \Delta_i =d$.
Since the integral is convergent in $d=6$, we can evaluate it there, ensuring that the latter relation is valid. 
This gives  the leading   term in $\epsilon = 6-d$, 
\be
 C = g_* + \ca O(\epsilon^{3/2}).
\ee
Using \Eq{FP-space} yields
\be\label{C-RG}
  {C = \sqrt{  \frac{6-d}{\frac5Q-\frac32}} + \ca O(\epsilon^{3/2})}.
\ee
We give the three most interesting values
\bea
C\Big|_{Q=1} &=&  \sqrt{   \frac{2}{7}(6-d) } + \ca O(\epsilon^{3/2}), \\
C\Big|_{Q=2} &=&  \sqrt{   {6-d} } + \ca O(\epsilon^{3/2}), \\
 C\Big|_{\rm Lee-Yang}&=&   C\Big|_{Q\to \infty} = i \sqrt{  \frac23 ({6-d}) } + \ca O(\epsilon^{3/2}).
 \label{C-LY}
\eea
Note that  for the Lee-Yang  class,  the definition involves an imaginary coupling, so this agrees with \cite{Goncalves2018}. 
In dimension $d=2$ this question has been solved analytically by relating \cite{DelfinoViti2010} the structure constant to the so-called DOZZ formula of imaginary Liouville conformal field theory \cite{Zamolodchikov2005},
a result which has been checked numerically \cite{PiccoSantachiaraVitiDelfino2013} and also generalised to a larger class of operators \cite{IkhlefJacobsenSaleur2016}. The structure constant is expressed in terms of the Barnes double gamma
function, whose evaluations for integer $Q$ read \cite{PiccoSantachiaraVitiDelfino2013}
\be
C(Q=1) = 1.0220, \quad C(Q=2) =1.0524, \quad C(Q=3) = 1.0923.
\ee
Our result is astonishingly precise for percolation in $d=2$. We will see that already for the Ising model this expansion can no longer be used below dimension $d=4$, see    section \ref{NPRG-Ising}. 
Finally, it is curious  that all known values of $C$ lie close to $C=1$, a value natural in dimensions $d=0$ and $d=1$, see appendix \ref{C-d=0+1}.

\subsection{The upper boundary of the non-critical domain} 
\label{The upper boundary of the non-critical domain}
\begin{figure}
\centerline{\fig{0.7}{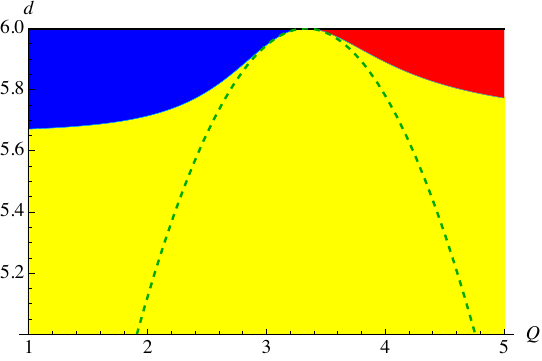}}
\caption{The critical line bounding the second-order phase at 2-loop order. In blue: $\lambda_{\rm c}>0$, in red $\lambda_{\rm c}<0$. In the yellow region there is a pair of complex conjugate fixed-points $\lambda_{\rm c} = \Re (\lambda_{\rm c})  \pm i \Im (\lambda_{\rm c}) $}
\label{f:QcDc}
\end{figure}
The $\beta$-function up to 2-loop order \cite{AlcantaraBonfirmKirkhamMcKane1981,KompanietsPikelnerPrivate}, divided by the coupling to exclude the Gaussian fixed point, reads  after some rewriting 
\be\label{B(u)}
B(u):=\frac{\beta(g)}{g}\Big|_{g^2=u} = \epsilon +\left(\frac{3}{2}-\frac{5}{Q}\right)
   u+\frac{Q (125 Q-794)+1340 }{72
   Q^2} u^2 +  \ca O(u^3)
\ee
The necessary condition for having a non-Gaussian fixed point is $B(u)=0$. 
As any quadratic equation, 
it has two solutions, of the schematic form $u_{1,2}= a \pm \sqrt b$. The solution relevant for us is the one which vanishes for $\epsilon\to 0$. As Fig.~\ref{f:QcDc} shows, as a function of $Q$ and $d$, there is a domain with one real positive solution (in blue), a domain with one negative real solution (in red), and a domain where no real solution exists, but a pair of complex conjugate ones (in yellow). The boundary is given by the line where $b$ vanishes. 
To leading order in $Q-Q_{\rm c}$, this reads
\be
d_{\rm c} = 6-\frac{729}{1480}(Q-Q_{\rm c})^2 + ...
\ee
This is the green dashed line in Fig.~\ref{f:QcDc}.

\begin{figure}[t]
\centerline{\fig{.55}{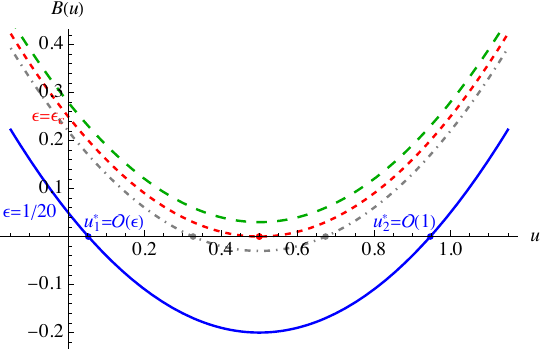}}
\caption{The function $B(u)$ defined in \Eq{B(u)} for varying values of $\epsilon$. For $\epsilon\ll 1$ there is a perturbative attractive solution $u_1^*=\ca O(\epsilon)$, and a non-perturbative repulsive (tricritical) solution $u_2^*=\ca O(1)$ (blue). Increasing $\epsilon$ the two solutions approach (gray, dot dashed) until they merge at $\epsilon=\epsilon_{\rm c}$ (red, dashed), and no solution exists for $\epsilon>\epsilon_{\rm c}$ (green, long dashes).}
\label{f:B(u)}
\end{figure}
To go beyond leading order, we need a more systematic procedure. 
Consider Fig.~\ref{f:B(u)}, where for visualization  we plotted  $B(u) = \epsilon -u +u^2$.   The coefficients   have the same signs as in \Eq{B(u)}, and the qualitative analysis for \Eq{B(u)} is the same.
 One sees that for $\epsilon$ small, there is a perturbative solution $u^*_1=\ca O(\epsilon)$, and a non-perturbative solution with $u_2^*=\ca O(1)$. The minimum of $B(u)$ is between these two solutions, and $B(u)$ is negative there. Up to $\epsilon=\epsilon_{\rm c}$ ($\epsilon_{\rm c}=1/4$ in the plot), there is still a solution for which both $B(u^*)=B'(u^*)=0$. For larger values of $\epsilon>\epsilon_{\rm c}$, no real solution exists, but a pair of   complex conjugate solutions. 
 
Our strategy to continue is   clear: We demand that 
\be
B(u^*) = B'(u^*) =0.
\ee
It is convenient to first write the latter equation, 
\be
0= B'(u) = \left(\frac{3}{2}-\frac{5}{Q}\right)
   +\frac{Q (125 Q-794)+1340 }{36
   Q^2} u +  \ca O (u^2 ). 
\ee
This equation is independent of $\epsilon$, i.e.\ dimension. 
It is   solved by  making for  $u^*$ an ansatz as a power series in $Q-Q_{\rm c}$.
Asking that $B(u^*)=0$   than gives $d_{\rm c}$ as a function of $Q-\frac{10}3$.    Using the 5-loop series   of \cite{KompanietsPikelnerPrivate}, partially given in  \cite{KompanietsPikelner2021,BorinskyGraceyKompanietsSchnetz2021}, to 3 loops in \cite{AlcantaraBonfirmKirkhamMcKane1981}, and to 4 loops in \cite{Gracey2015}\footnote{Due to undefined objects  in the auxiliary  file provided in \cite{Gracey2015} we were unable to check these results against \cite{KompanietsPikelnerPrivate}.} we get
\bea\label{dcofQ5loop}
d_{\rm c} &=& 6-0.492568  \textstyle \left(Q- \frac{10}{3}\right)^2-1.49158
   \left(Q-\frac{10}{3}\right)^3-14.9483 \left(Q-\frac{10}{3}\right)^4 \nn\\
&&\textstyle -184.253
   \left(Q-\frac{10}{3}\right)^5  +\ca O \left(Q-\frac{10}{3}\right)^6.
\eea
This series is a perturbatively controlled series, and seemingly Borel-summable for $Q<10/3$:  having $Q_c{-}Q\sim \sqrt \epsilon$ makes the coupling $u\sim \sqrt \epsilon$, and the terms dropped in \Eq{B(u)} of order $\epsilon^{3/2}$.
 The results for the Borel summation of \Eq{dcofQ5loop} are given in Fig.~\ref{f:NPRG-phase-diagram} \cite{KompanietsPrivate,KompanietsWieseToBePublished}.
For $Q>Q_{\rm c}=10/3$, all terms of the series are negative, which indicates that a branch cut singularity starts there. 
We come back to this question in our NPRG treatment in section \ref{s:upper-right-branch}.

 We finally  note that cubic theories were also proposed  for $SU(N)$ \cite{McKaneWallaceZia1976,Gracey2015,GraceyRyttovShrock2020},  $O(N)$ broken by an additional cubic interaction \cite{FeiGiombiKlebanovTarnopolsky2015}, or $O(N)\times O(m)$ \cite{GraceySimms2017}. A similar analysis can be performed there.

\begin{figure}[t]
\fig{0.49}{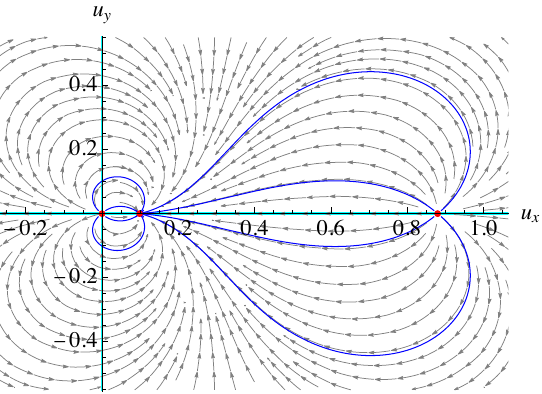}\hfill\fig{0.49}{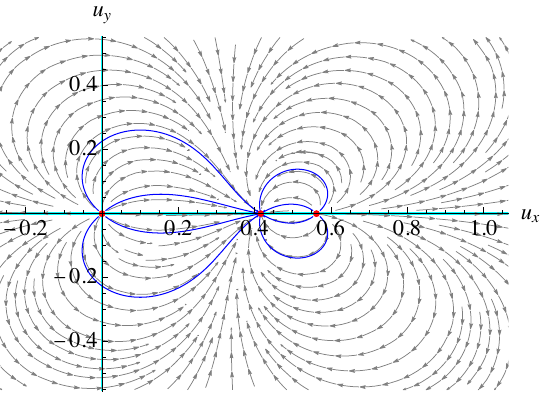}\\
\centerline{$\epsilon=0.01$ \hspace{0.45\textwidth}$\epsilon=0.0027$}
\fig{0.49}{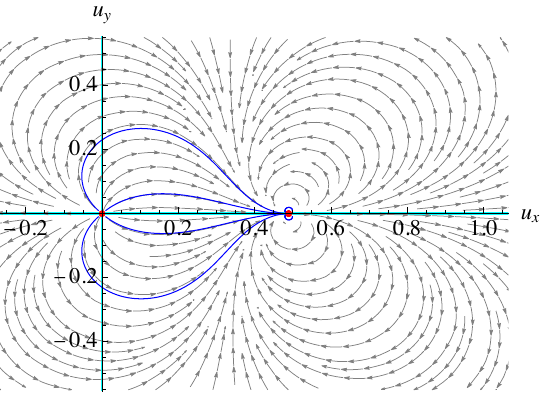} \fig{0.49}{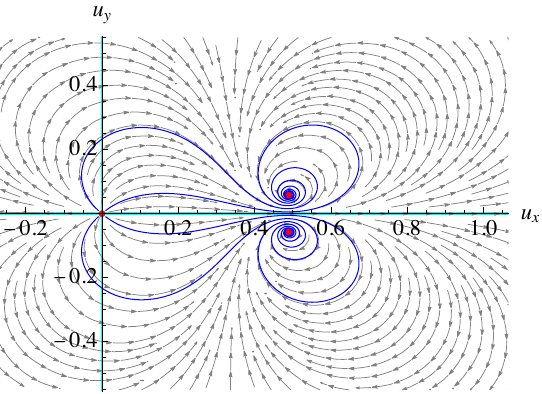}
\centerline{$\epsilon=0.027613$ \hspace{0.4\textwidth}  $\epsilon=0.0028$~~~}
\caption{The flow-diagram for the $Q$-state Potts model at 2-loop order in minimal subtraction at $Q=3.1$,  $\epsilon=0.01$ (top left), $\epsilon=0.0027$ (top right), 
$\epsilon=0.027613$ (lower left) and $\epsilon=0.028$ (lower right). In blue are RG trajectories starting close to the fixed points.
On the first two graphs, the   leftmost Gaussian FP has two repulsive directions, the non-trivial critical fixed point in the middle has two attractive directions, whereas the tricritical FP to the right has two repulsive directions. While the Gaussian fixed point remains completely repulsive, the pair of complex-conjugate FPs have a complex eigenvalue with positive real part. As a result, all trajectories spiral in, except for the real axis ($u_y=0$), which runs to strong coupling $u\to \infty$. 
}
\label{f:RG-flow-complex-plane}
\end{figure}

It is interesting to see what happens to the RG flow in the complex plane. This is done on Fig.~\ref{f:RG-flow-complex-plane}. For small $\epsilon$ (here $\epsilon=0.01$, top left plot) the critical fixed point lies close to the Gaussian one, while a tricritical (bi-unstable) fixed point is at $u\approx 0.9$. Increasing $\epsilon$ (top right), the critical and tricritical fixed points approach,  until they merge at $\epsilon = \epsilon_{\rm c}=0.0027$ (lower left plot). Up to this value of $\epsilon$, all fixed points lie on the real axes $u_y=0$. Increasing $\epsilon$ further, a pair of complex-conjugate fixed points emerges. 
Since the critical fixed point for smaller $\epsilon$ is globally attractive and the flow at large couplings is not rearranged, the pair of complex-conjugate fixed points is also globally attractive, with the RG flow spiraling in (complex eigenvalues). 

\section{Non-perturbative renormalization}
\label{s:Non-perturbative renormalization}
\subsection{Flow equations}
Mapping out the full phase diagram is impossible by relying solely on controlled expansions. Here we study the full phase diagram 
using the different non-perturbative RG schemes, LPA, LPA${}'$, LPA${}^*$ and  Wilson's original approach. 
For the Potts model, this line of research  was pioneered in \cite{NewmanRiedelMuto1984} (Wilson scheme), and continued in \cite{ZinatiCodello2018,Sanchez-VillalobosDelamotteWschebor2023} (LPA${}'$). 
The   idea is to start from the action
\be
\ca S[\Phi] = \sum_\alpha \half \left[\nabla \Phi^\alpha(x) \right]^2 + \frac12 U(\Phi), 
\ee
where\footnote{In most of the current literature \cite{ZinatiCodello2018,Sanchez-VillalobosDelamotteWschebor2023} the expansion is written in terms of the {\em unconstrained basis}, see appendix \ref{a:algebraic-objects}. This is much more tedious to implement than in the {\em constrained basis} used here.}
\bea\label{U(Phi)}
U(\Phi) &=& \sum_{ \alpha=1}^Q   \lambda_2 (\Phi^\alpha)^2+ \lambda_3 (\Phi^\alpha)^3  + \lambda_4 (\Phi^\alpha)^4  + \lambda_5 (\Phi^\alpha)^5  + \lambda_6 (\Phi^\alpha)^6 + ...\nn\\
&+& \sum_{ \alpha=1}^Q \sum_{\beta=1}^Q \lambda_{22}   (\Phi^\alpha)^2 (\Phi^\beta)^2 + \lambda_{23}   (\Phi^\alpha)^2 (\Phi^\beta)^3 +  \lambda_{24}     (\Phi^\alpha)^2 (\Phi^\beta)^4 
+ \lambda_{33}   (\Phi^\alpha)^3 (\Phi^\beta)^3 +   ... \nn\\
&+& \sum_{ \alpha=1}^Q \sum_{\beta=1}^Q   \sum_{ \gamma=1}^Q   \lambda_{222}    (\Phi^\alpha)^2 (\Phi^\beta)^2  (\Phi^\gamma)^2 + ...
\eea
The correction to the effective action  in the Wilson scheme  is obtained by integrating out the largest wave-vector mode, here with $\Lambda\to 1$. (The index structure is given later.)
\bea
\frac{\delta U(\Phi)}{2} &=&   \delta \ca S[\Phi]  =     \ln \left( \ca S''[\Phi]\right)+ \mbox{const} =    \ln \left(\frac{\ca S''[\Phi]}{\ca S''[0]}\right)
=      \ln \left(\frac {1+\half U''(\Phi)}{1+\half U''(0)}\right) \nn\\
&=&     \ln \left(1 +\frac {\half[U''(\Phi)-U''(0)]}{1+\half U''(0)}\right) =   \sum_{n=1}^\infty \frac{(-1)^{n+1}}n \left(\frac {U''(\Phi)-U''(0)}{2(1+\lambda_2)}\right)^n .
\eea
This leads to 
\be\label{106}
{\delta U(\Phi) = \sum_{n=1}^\infty \frac{\left(-\half \right)^{n-1}}n \left(\frac {U''(\Phi)-U''(0)}{ 1+\lambda_2 }\right)^n}.
\ee 
In contrast, in NPRG one has 
\bea\label{NPRG1}
\frac{\delta U(\Phi)}{2} &=&  \delta \ca S[\Phi]  =   \frac {-1}{1+\half U''(\Phi)} = \frac {-1}{[1+\half U''(0)] + \half [U''(\Phi)-U''(0)]} \nn\\
&=& \sum_{n=1}^{\infty}   {\left(-\half \right)^{n}} \frac { \left[ U''(\Phi)-U''(0)\right]^n}{ (1+\lambda_2) ^{n+1}}.
\eea
This gives
\be
{\delta U(\Phi) = \sum_{n=1}^\infty  {\left(-\half \right)^{n-1}}   \frac {[U''(\Phi)-U''(0)]^n}{ (1+\lambda_2)^{n+1} }}.
\ee
Note the power of $1+\lambda_2$ in the denominator, which is larger by one than in Wilson. 
The difference in combinatorial factor can be rationalized as follows: The Wilson cutoff is a hard cutoff, which allows one to   integrate out the fastest mode, leading to $\ln (\ca S''[\Phi])$. 
The   cutoff used for the LPA is a soft cutoff, the calculatorially easiest choice is the Litim cutoff \cite{Litim2001}
\be
R_{k} (p ) = (k^2-p^2) \Theta (|p|<k). 
\ee
With this choice the IR flow reads
\bea
- k \partial _k S[\Phi] &= &- \frac12 k \partial _k\int_p \ln \left(p^2+ \half U''(\Phi) + R_k(p)\right) =-\frac12  \int_p \frac{k \partial_k R_k(p) } {p^2 + \half U''(\phi)+ R_k(p)} \nn\\
& = & - k^2 \int_p \frac{\Theta(|p|<k) } {k^2 + \half U''(\phi)} = \frac{S_d}{d(2\pi)^{d}}    \frac{-k^{d+2} } {k^{2} + \half U''(\phi)}. 
\label{NPRG2}
\eea
Scaling $k\to 1$ and absorbing the volume factor gives \Eq{NPRG1}. 

Finally, we need to rescale $U$ and $\Phi$, and add indices. This leads to the NPRG equation (IR flow)
\be\label{beta}
 \partial_ \ell U(\Phi) = d U(\Phi) - \frac{d-2+\eta}2 \sum_\alpha \Phi_\alpha \frac{\partial U(\Phi)}{\partial \phi_\alpha} + \sum_{n=1}^{\infty}   \ca C^n_{\lambda_2} {\left(-\half \right)^{n-1}}  \mbox{tr} \left(\frac { [ \mathbb U''(\Phi) {-}\mathbb U''(0)] \cdot\mathbb P }{1+\lambda_2}\right)^n .
\ee
The first remark is that a global prefactor (the normalization of space) can   be absorbed by a rescaling of $U$, as this changes the rescaling terms which are linear in $U(\Phi)$, but not $\delta U(\Phi) $. 

We have  written explicitly the index structure of each term in  the Potts model, with the matrix $\mathbb U'' (\Phi)$ and the projector $\mathbb P$ (with the same index structure as the propagator) defined as (see appendix \ref{a:algebraic-objects})
\be
\mathbb U'' (\Phi)_{\alpha\beta} = \frac{\partial^2 U(\Phi) }{\partial \Phi_\alpha \partial \Phi_\beta}, 
\qquad \mathbb P_{\alpha \beta} :=   \vec e_{\a} \cdot \vec e_\beta = \delta _{\alpha \beta}-\frac1Q.
\ee
\begin{figure}[t]
\fboxsep0mm
{\setlength{\unitlength}{1mm}
\begin{picture}(169,105)
\put(0,0){\fig{0.6}{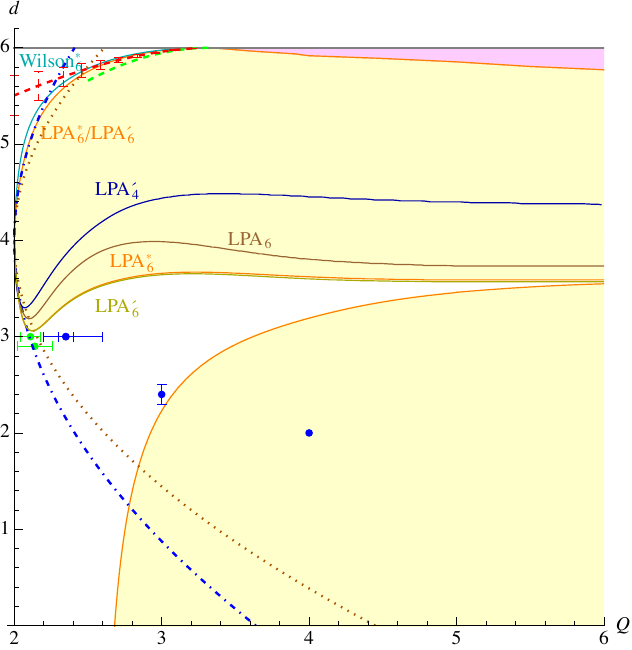}}
\put(110,0.5){\fig{0.35}{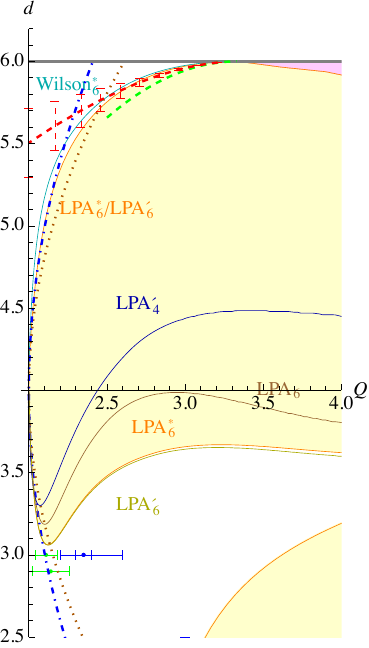}}
\end{picture}}
\caption{Left: The phase diagram of Fig.~\ref{f:Potts-phase-diagram} as given by LPA${^*_6}$ (yellow, with orange borders), almost indistinguishable from LPA${'_6}$ (olive), and compared to LPA${_6}$ (brown) and LPA${'_4}$ (dark blue). The blue dot-dashed line is the expansion \eq{curvatureNPRG}; dark red dotted is for \Eq{NewmanRiedelMuto1984}. For the Wilson scheme we show the boundary in the upper left quadrant (cyan). 
The green dashed line is the leading order of \Eq{dcofQ5loop}, while the red  dashed line is a weighted average of Pad\'e and Pad\'e-Borel resummations of the 5-loop result \cite{KompanietsPrivate,KompanietsWieseToBePublished}.
The blue dots are results obtained by other methods, see Fig.~\ref{f:Potts-phase-diagram}. Right: Blow up of upper left quadrant.}
\label{f:NPRG-phase-diagram}
\end{figure}%
The factor of $\ca C^n_{\lambda_2}$ depends on the cutoff and $\lambda_2$, and reads
\be
\ca C^n_{\lambda_2} = \left \{{ \begin{array}{cc} \frac{1}{1+\lambda_2}\rule[-4mm]{0mm}{3mm} & \mbox{~~Litim-cutoff (LPA)~~}\\
n^{-1} & \mbox{~~hard cutoff (Wilson)~~} 
\end{array}} \right. .
\ee
We   added an   anomalous dimension for the field, or exponent  $\eta$.
The latter is obtained from 
\be\label{eta}
\eta = \frac{\ca C_\eta}{2(1+\lambda_2)^4} \mbox {tr}  \left( \frac{\partial \mathbb U''(\Phi) } {\delta \Phi_\alpha }\cdot \mathbb P  \cdot\frac{\partial \mathbb U''(\Phi) }{\delta \Phi_\alpha } \cdot \mathbb P \right) \bigg|_{\Phi=0}   = \frac{9  {\ca C_\eta} \lambda_3^2 }{(1+\lambda_2)^4  }\frac{Q-2}{Q}. 
\ee
The global prefactor $\ca C_\eta$ is fixed s.t.\ for $d\to 6$ the $\beta$-function vanishes at $Q=10/3$, 
\be
\ca C_\eta = \left \{{ \begin{array}{cc} 1 & \mbox{~~Litim-cutoff (LPA)~~}\\
\frac13 & \mbox{~~hard cutoff (Wilson)~~} 
\end{array}} \right. .
\ee
This is equivalent\protect\footnote{For the Wilson scheme and $\lambda_2\to 0$ this reduces to \Eq{eta(g)} noting that $\lambda_3 = g/3$, and that there is an additional explicit factor of $2$ in \Eq{106} as compared to the field theory.} to \Eq{eta(g)} for $\eta(g)$.
When improving LPA, we call this scheme LPA{${}^*$}, when improving Wilson we call it  Wilson{${}^*$}.
It is slightly different from what is   used in the NPRG literature \cite{Sanchez-VillalobosDelamotteWschebor2023}, and termed $\mbox{LPA${}'$}$: There is an additional factor of $(1-\frac{\eta}{d+2})$ multiplying the r.h.s.\ of \Eq{NPRG2}.  
The reader may wonder about the denominator $1/(1+\lambda_2)^4$ in \Eq{eta}. 
A heuristic way to derive this is to take one derivative w.r.t.\ the passing momentum $p^2$, which is similar to a variation of the squared mass $\lambda_2$, increasing the number of factors in the denominator by 1. 

All these subtleties are unimportant close to the upper critical dimension 6, and for lower dimensions everything we do here seems badly  controlled: This is seen when comparing the different approximation schemes in order to see how robust the results are. This is done later; we will see that   LPA${}'$ and LPA${}^{*}$ (at the same field order) are almost indistinguishable, but that for small $d$ the results strongly depend on the maximal number of fields allowed.

\subsection{Implementation}
We implement the above program by going up to order 6 in the fields, as suggested by \Eq{U(Phi)}. 
We then calculate the   corrections to $U(\Phi)$ up to order $U^7$. This perturbative treatment, also used in \cite{ZinatiCodello2018,Sanchez-VillalobosDelamotteWschebor2023}   properly accounts for the 
algebraic structure of the Potts propagator\footnote{This is possible at fixed $Q$, when restricting to the spin degrees of freedom. As an example, for $Q=3$, there are two independent fields, and the theory  can be written in terms of $\sum_\alpha \Phi_\alpha^2$ and $\sum_\alpha \Phi_\alpha^3$. For non-integer $Q$ this is not possible.}. We do not know of a truly non-perturbative approach at non-integer $Q$, where $U''(\Phi)$ as a function appears in the 
denominator.  

We then 
start somewhere below dimension $d=6$, 
propose an initial condition with $\lambda_3\neq 0$, and try to evolve to a fixed point of the $\beta$-function \eq{beta}.
We have several routines to do so: The first tries to integrate the flow-equation itself, in which we have {\em reversed} the flow for $\lambda_2$, which is a relevant coupling: it has an RG eigenvalue close to 2. The other algorithms we implemented are different  routines to  find a nearby minimum of $|\beta|$
(steepest descent, Newton iteration, Monte Carlo). 
Once we find a solution, we can walk in the $\{ Q,d \}$ plane, following this solution under a change of $Q$ or $d$,  until we reach the critical line. What happens there depends on where we start. Let us discuss the different sectors. If not stated differently, we use the LPA${{}^*}$ scheme where $U$ is truncated at   order $\Phi^6$, and which we denote LPA${{}^*_6}$. Results for LPA${{}'_6}$ are almost indistinguishable. 

\subsection{Branches}
\subsubsection{Upper left sector ($Q<10/3, 4<d<6$)} For $2<Q<10/3$ fixed and $d=5.99$, or $d=5.999$ (close to $Q=10/3$)   we propose a solution with $\lambda_3=0.2$, and the remaining $\lambda_i=0$. Using the $\beta$-function where the flow of $\lambda_2$ is reversed, these couplings  converge to a critical point $\vec \beta [\lambda_i]=0$ of the $\beta$-function. Linearizing the $\beta$-function around this solution we find one relevant EV, while all remaining EVs are irrelevant. We then decrease $d$, and follow the fixed point in question by again minimizing the $\beta$-function. At some point,  the second-largest EV becomes  zero, at which point it merges with a tricritical fixed point. Further decreasing $d$ we   loose this fixed point and as a consequence run to strong coupling; in practice, our algorithm no longer succeeds to finds a  solution, and   blows up.  We identify this dimension as  $d_{\rm c}(Q)$.  If we worked hard, we should be able to find a tricritical point, which merges with the former fixed point at $d_{\rm c}(Q)$.  One observes that close to $d_{\rm c}(Q)$ this eigenvalue behaves as a constant times $\sqrt{d-d_{\rm c}(Q)}$. 
Close to $d=4$, $Q=2$ the curvature of this curve is close to what is expected within the NPRG. 
We had expected to see the expansion of Ref.~\cite{NewmanRiedelMuto1984} given in \Eq{NewmanRiedelMuto1984} to agree with our Wilson$_6^*$ approximation, 
but this is not the case. As we do not have the equations of \cite{NewmanRiedelMuto1984} to compare with, we cannot make this comparison quantitative. 

Finally, we could try to follow the fixed points into the complex plane. We did not pursue this approach here since we would have to double the number of couplings   passed to our routines.

\begin{figure}
\centerline{\fig{0.6}{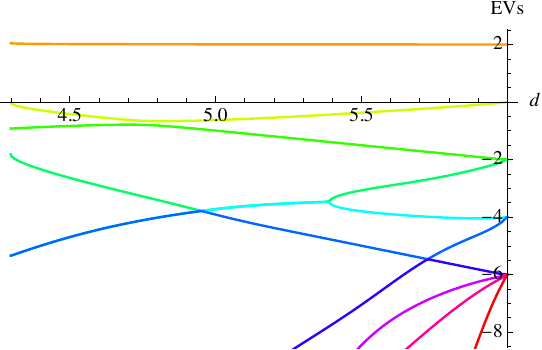}\hfill\fig{0.371}{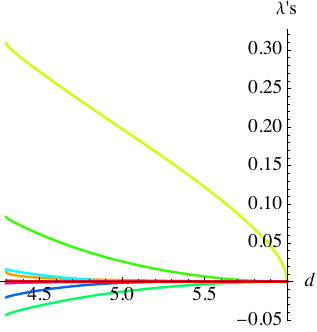}}
\caption{Left: Real part of   eigenvalues for the $\beta$-function in varying dimension $d$, for $Q=2.01$. The program starts in $d=5.99$, where one sees the operator content of the free theory: one relevant eigenvalue 2 for $\lambda_2$, the marginal eigenvalue for $\lambda_3$, two quartic operators of dimension $-2$, two
quintic operators of dimension $-4$, and four order-6 operators. Arriving in dimension $d=4.2975$, one eigenvalue becomes marginal and we loose the critical fixed point. When two lines meet, they form a complex conjugate pair, and we only see a single line continuing. Right: the coupling constants. Nothing special seems to happen at $d_{\rm c}(Q)$.}
\label{f:EVs}
\end{figure}

\subsubsection{Upper right sector  ($Q>10/3, 4<d<6$)}
\label{s:upper-right-branch} 
\begin{figure}
\centerline{\fig{0.5}{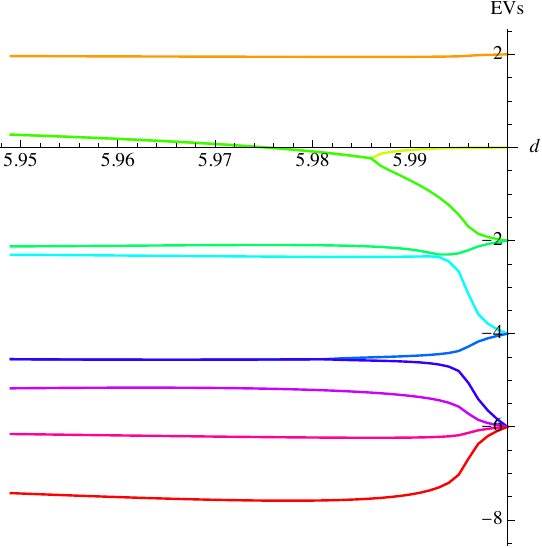}\hfill\fig{0.5}{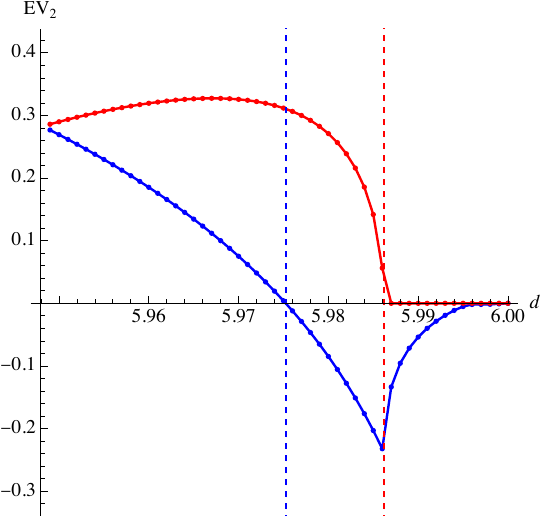}}
\caption{Left: real part of eigenvalues for the $\beta$-function in varying dimension $d$, for $Q=3.45$. The program starts in $d=5.999$, where one sees the operator content of the free theory. Then the program     descends to  $d=5.95$,  well below the upper critical dimension, defined by the dimension where the real part of the subleading EV becomes positive. On the right we show the real part (in blue) and the imaginary part (in red) of the second-largest EV. The dashed line denotes the dimensions $d_1$ where the imaginary part starts to be non-vanishing, and $d_2$ where the real part becomes positive.} 
\label{f:EVs2}
\end{figure}

The scenario observed for $Q<10/3$ is also observed for  $Q>10/3$, after multiplying the odd couplings by a factor of $i$. However, close to $Q_{\rm c}$, we see a different behavior, for which an example at   $Q=3.45$ is shown in Fig.~\ref{f:EVs2}. What we observe is that at a dimension $d_1$ the third-largest EV merges with the second largest one, and together they wander off in the complex plane. Decreasing $d$ further to $d_2$,  they both become relevant, albeit with 
a non-vanishing imaginary part. Below the latter dimension, the RG equation runs to strong coupling, so $d_{\rm c}\ge d_2$. We suspect, however, that the critical dimension is already reached at $d_1$; otherwise the   critical dimension as a function of $Q$ seemingly jumps at $Q=10/3$, which we have a hard time believing. On the other hand,  up to 5-loop order the perturbative series for $d_{\rm c}(Q)$ has only positive terms for $Q>Q_{\rm c}=10/3$, whereas the series alternates for $Q<Q_{\rm c}$. This may signal a non-analytic behavior for $d(Q)$ for $Q>Q_{\rm c}$. Our results indeed show a roughly linear dependence of $d_{\rm c}(Q)$, on 
$ Q-Q_{\rm c}$. 

It is equally possible  that the   approximations  used, either in the NPRG or its implementation, are inadequate. 
There should not be a problem with the numerical evaluation of the zeros of the $\beta$-function, which is done with 400 digits of precision (more than 10 times machine precision). 

We finally note that within the NPRG in the limit of $Q\to \infty$ the $Q$ copies not necessarily decouple, as it is possible to have a fixed point with an inter-copy  coupling (as $\lambda_{23}$) persist when taking $Q\to \infty$, even though when starting with decoupled models at $Q=\infty$, these inter-copy interactions are not generated. 
As a result, this limit may be  as subtle as the large-$N$ limit for Random-Field \cite{MezardYoung1992,LeDoussalWiese2005b}.

\subsubsection{Lower  branches}
The lower left branch is   more difficult to reach, as generically most initial conditions  either blow up when evolving them with the  ``improved'' $\beta$-function, or converge to multiply unstable (tricritical or higher) fixed points. We also tried to walk down for $Q<2$, which is possible, but then we failed to cross the line $Q=2$, where some of the operators decouple. 
Our successful strategy   was to generate random couplings $0<\lambda_3<1$,  $-0.5<\lambda_4<0.5$, and $-0.5<\lambda_{22}<0.5$, with $\lambda_i=0$ for the rest. We then used steepest descent for  $Q=2.5$ and $d=2.2$ towards a fixed point, and stop when this fixed point has  one relevant direction. The successful trial for LPA${_6^*}$ converged to
\bea
&&\lambda_2 = -0.331907, \lambda_3 = 
  0.139233, \lambda_4 = 
  0.0481672, \lambda_{22} = 
  0.00200003, \lambda_{5} = 
  0.0119803, \nn\\
  &&\lambda_{23} = -0.00104314, \
\lambda_6 = 0.00399691, \lambda_{24} = 
  0.000456486, \lambda_{33} = -0.000520252, \nn\\
&&\lambda_{222} = 0.00010586.
\eea
To proceed, we start at this point,  and then walk in the $(Q,d)$ plane  until we hit the boundary of the critical region. 
The result is shown in Fig.~\ref{f:NPRG-phase-diagram}. It is reassuring to see that the various LPA approximations all have the same parabolic shape around $d=4$, $Q=2$. Depending on which approximation is used, the critical line follows this parabola   further, or less. The results of 
Ref.~\cite{Sanchez-VillalobosDelamotteWschebor2023} (green points with error bars closes to $d=3$), which go up to  order 9 in the field, are able to follow this parabola a little further than we do, without noticeably deviating from it.  
When comparing the different schemes, one   observes that all schemes follow the critical parabola 
given in \Eq{curvatureNPRG} for $d<4$, and   that   including more fields allows one to follow this parabola further. 
 Including $\eta$ also increases the range for which this is possible, i.e.\ before $d_{\rm c}(Q)$ grows again. 
There is only a minimal difference between LPA${}'$ and LPA${^*}$. 
We were unable to find a value of the couplings to repeat this analysis in the Wilson scheme. 
We conjecture that this parabola is close to the critical curve, and a slight deformation to pass through 
$d=2$, $Q=4$ is a good  approximation for the true critical curve. This is the rational behind   Fig.~\ref{f:Potts-phase-diagram}.

Going back to Fig.~\ref{f:NPRG-phase-diagram}, we observe that 
the critical curve $d_{\rm c}(Q)$ becomes horizontal for large $Q$, and by walking down we find a second curve 
as indicated in Fig.~\ref{f:NPRG-phase-diagram}. While this may well be an artifact of the scheme, it leaves open the intriguing possibility that in 
dimension $d=3$ the  Potts model has a window of values for $Q$ for which it becomes critical again.

\begin{figure}[t]
\centerline{\fig{0.61}{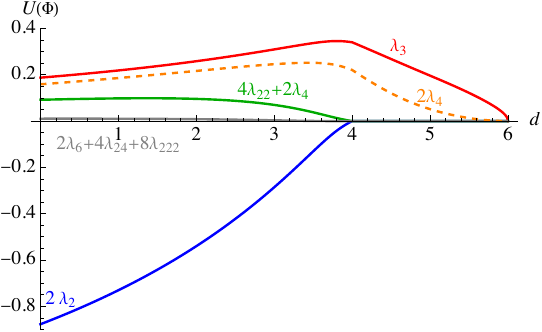}\hfill\fig{0.34}{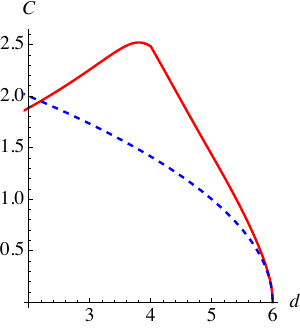}}
\caption{Left: The couplings for the Ising model as a function of $d$. Right: The structure constant $  C$ from the NPRG (red, solid), as compared to perturbation theory (blue dashed). The non-analytic behavior at $d=4$ is clearly visible, thus field theory in dimension $d=6-\epsilon$ will stop to work in dimension $d=4$.}
\label{f:Ising-walk}
\end{figure}
\begin{figure}[t]
\centerline{\fig{0.6}{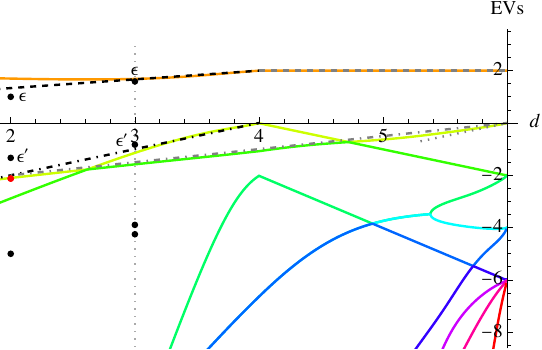}}
\caption{Eigenvalues of the NPRG $\beta$ function for the Ising model. The largest eigenvalue is $1/\nu$, which equals 2 in dimensions $4\le d\le 6$, and has $\epsilon$-expansion $1/\nu=2-(4-d)/3$ below dimension $d=4$. The exponent $\omega$
is the second largest eigenvalue. For the   cubic coupling it has 1-loop dimension $-\omega=d-6$ in   $\epsilon=6-d$ (gray dotted), while $(d-6)/2$ (gray dot-dashed) seems to be a decent approximation   down to $d= 2$. The quartic coupling has 1-loop eigenvalue $-\omega = d-4$ (black dashed). 
The NPRG flow respects   these expansions. The black dots in $d=3$ show the location of  scalar even-spin operators ($\epsilon$, $\epsilon'$...) in the numerical bootstrap, see table 2 of  \cite{Simmons-Duffin2017}. They do not contain the eigenvalue for the cubic coupling.
In $d=2$ we use that the $\mbox{EVs}=2-2 h_{r,s}$, with $h_{r,s}= [(4 r-3 s)^2-1 ]/48$, $r=2$, $s=1$ for $\epsilon$, and $r=3$, $s=1$ for $\epsilon'$; we also added $r=4$, $s=1$. We conjecture that the leading odd operator with coupling $\lambda_3$ evolves to the first irrelevant operator in the magnetic series \cite{JacobsenSaleur2019} $2-2 h_{1/2+n,0}$, $n \in \mathbb N_0$, which gives $2-2 h_{5/2,0}= -17/8$, at $n=2$ (red dot).}
\label{EVs-Ising-NPRG}
\end{figure}

\subsection{Ising ($Q=2$)}
\label{NPRG-Ising}
We succeeded to walk down at $Q=2$ from $d=6$ to $d=0$.  The result is shown on the left of Fig.~\ref{f:Ising-walk}, using LPA: since $
\eta\sim Q-2=0$, the schemes are identical, i.e.\ $\mbox{LPA}=\mbox{LPA}'=\mbox{LPA}^*$. 
We can then project onto the Ising spin variables, by writing down $U(\Phi)$ in terms of $\Phi^1$ and $\Phi^2$,   then setting $\Phi^1\to \phi$, $\Phi^2\to -\phi$.
This gives the potential visible in the spin sector of the Ising model.
\be\label{u(phi)}
U(\Phi) \to u(\phi)= 2 \lambda _2 \phi
   ^2 +2 \left(\lambda _4+2
   \lambda _{22}\right) \phi ^4 + 2 \left(\lambda _6+2 \lambda _{24}+4 \lambda
   _{222}\right) \phi ^6 + ....
\ee
As can be seen in Fig.~\ref{f:Ising-walk}, these couplings  vanish in dimensions $4\le d\le 6$, even though $\lambda_4$ itself is non-vanishing there. This is  the Gaussian fixed point for the spin-degrees of freedom in dimension $d>4$.
Below dimension $d=4$, the quartic coupling visible in the spin theory, see \Eq{u(phi)}, becomes non-vanishing. Descending below $d\approx 3$ also the sextic couplings become visible, albeit small. 

This plot leads to the conjecture that the structure constant $  C$, at leading order proportional to $\lambda_3$, grows from dimension $d=6$ to dimension $d=4$, before descending again. Assuming that the structure constant $\ca C$ is indeed proportional to $\lambda_3$, and normalizing s.t.\ the $\epsilon$-expansion is matched, gives the plot on the right of Fig.~\ref{f:Ising-walk}. It shows clearly that the $d=6-\epsilon$ expansion for the structure constant given in section \ref{3-point function and structure factor C} will break down in dimension $d=4$. 

There are some interesting tests we can perform, see Fig.~\ref{EVs-Ising-NPRG}: The largest eigenvalue in the stability matrix of the 
$\beta$-function is $1/\nu$, which equals 2 in dimensions $4\le d\le 6$, and has $\epsilon$-expansion $1/\nu=2-(4-d)/3$ below dimension $d=4$. This is well respected in the NPRG, see the topmost curve on Fig.~\ref{EVs-Ising-NPRG}.
The exponent $\omega$ is the second-largest eigenvalue. 
For the   cubic coupling it has 1-loop dimension $-\omega=d-6$ in   $\epsilon=6-d$ (gray dotted in Fig.~\ref{EVs-Ising-NPRG}) which is valid for $5\ll d <6$, while $(d-6)/2$ (gray dot-dashed in Fig.~\ref{EVs-Ising-NPRG}) seems to be a decent approximation   down to $d= 2$. The quartic coupling has 1-loop eigenvalue $-\omega = d-4$. 
The NPRG flow respects   these expansions. The operator content seems rather sparse in dimension 
$d=3$, and misses higher scalar operators known in the bootstrap \cite{Simmons-Duffin2017}. This should improve upon increasing the maximum field dimension in the NPRG. 

\begin{figure}
\fig{.6}{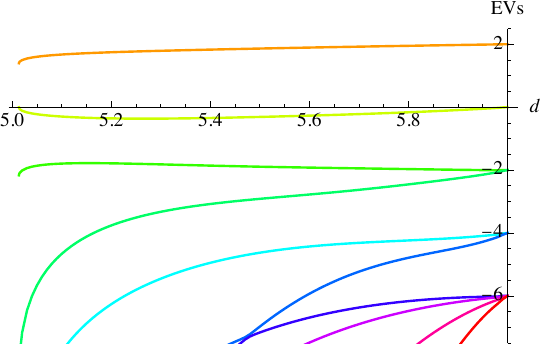}\hfill\fig{.388}{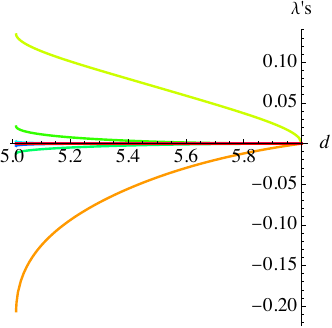}
\caption{Eigenvalues and couplings when walking down at $Q=1$ from $d=5.99$.}
\label{f:walkQ=1}
\end{figure}

\subsection{$Q<2$}
We can also walk down from $d=6$ for $Q<2$. For $Q=3/2$, this is possible down to dimension $d=0$. 
For $Q=1$ the system of RG equations seem to become degenerate in dimension $d=5.013$, see Fig.~\ref{f:walkQ=1}.
For $Q=1/2$ this happens already in dimension $d=5.251$. This may be   a sign for a first-order transition (which is unlikely given what we know about percolation), or a       technical problem. In the latter case it may indicate that the $\epsilon$-expansion is valid only down to dimension $d\approx 5$. Another degeneracy is observed when trying to walk in dimension $d<4$ from $Q<2$ to $Q>2$. 
It would be interesting to analyze this further. 

\section{Discussion and conclusion}
\label{s:Conclusion}
We showed that there are two distinct field theories for the $Q$-state Potts model, one for the spin degrees of freedom, and one for the cluster degrees of freedom. The latter is   exact on any graph, and does not impose the   representation of the $Q$ states as a vector in $\mathbb R^{Q-1}$. As a consequence of these   different representations, 
the Ising model has  two distinct 
upper critical dimensions,  $d_{\rm c}=4$ for spin observables, and $d_{\rm c}=6$ for cluster properties. 
This allowed us to derive  an explicit prediction for the 3-point function, equivalent to the structure constant of the 
underlying CFT. We hope that numerical simulations  will soon verify this prediction.

An interesting question is how this extends to other values of $Q$. Our analysis shows that for $Q>2$ the Potts model 
has a second-order transition  only outside a non-critical region. Its boundary is  perturbatively 
controlled close to $d=6$, with a parabola, open to the bottom in the $(Q,d)$ plane, emanating from  $Q=10/3$, $d=6$, see Fig.~\ref{f:Potts-phase-diagram}.
Another bounding parabola open to the right, emanates from $d=4$, $Q=2$, with a coefficient which can be estimated within a non-perturbative approach. As discussed in the introduction, there is ample numerical and analytical evidence that the lower bound of the non-critical region exits as well, even though it is yet badly approximated by any RG scheme. We are working on a $d=2+\epsilon$ expansion to remedy this. One should also be able to expand around the point $Q_{\rm c}=4$ and $d=2$, similar to what was done in \cite{CardyHamber1980} (up to an unknown coefficient) for the $O(N)$ model (around $N=d=2$), see also \cite{HaldarTavakolMaScaffidi2023}.
Going back to $d\approx 6$ and $Q=10/3$, we note that   
a new critical theory emerges when rotating all odd couplings into the complex plane. 
Taking $Q\to \infty$ this connects to the well-known Young-Lee universality class.

Since the clusters of the FK expansion live inside the spin clusters (see section \ref{s:Sampling cluster configurations from spin configurations}),  spin correlations cannot fall off faster than cluster correlations. Written for the field-dimension $\Delta_\phi$, or the exponent $\eta$, this reads
\be\label{cluster-inequality}
\Delta_{\phi}^{\rm spin} \le \Delta_{\phi}^{\rm cluster} \qquad \Leftrightarrow \qquad \eta^{\rm spin} \le \eta^{\rm cluster}.
\ee
\Eq{eta-cluster} shows that to 1-loop order, and for $d<6$
\be
\eta^{\rm cluster}\ge 0 \quad \mbox{for}\quad 2< Q < \frac{10}3.
\ee
This property persists to higher orders, and especially for $d$ close to $4$ ($\epsilon=6-d\approx 2$). 
For the Ising model $\eta^{\rm cluster}=\eta^{\rm spin}=0$ for $4\le d\le 6$, and the bound is saturated.

Perturbative RG near dimension 6 gives a   clear picture of what happens when we enter the ``first-order'' domain: allowing the couplings to become complex,   the critical and an additional tricritical point merge, and then wander into the complex plane, forming a pair of complex conjugate critical fixed points.  This complex CFT inside the ``first-order'' regime can be accessed by expanding around $d=6$ and $Q=10/3$ \cite{KompanietsWieseToBePublished}. In contrast,  when starting with real couplings, the RG flow runs to strong coupling, a sign (but no proof) that the phase transition in this domain is first order. 

The pair of complex fixed points inside the ``first-order'' regime  can be studied     
via numerical simulations in $d=3$, or via transfer matrix in $d=2$. 
Such a complex CFT   was   conjectured to exist  in $d=2$ for $Q>4$ 
\cite{GorbenkoRychkovZan2018b,MaHe2019}, and   realized in a lattice model  \cite{JacobsenWiese2024}.
The resulting   CFT has a complex central charge, and a complex spectrum, given by the proper analytic continuation of the CFT data for $Q<4$.

\subsection*{Acknowledgements}
We thank Mikhail Kompaniets and Andrey Pikelner for providing us with the 5-loop perturbative expansion for the $Q$-state Potts model, and John Cardy, Bernard Julia, Mehran Kardar, Adam Nahum and Slava Rychkov for stimulating discussions. 
We are most indebted to  Andrei Fedorenko  for generously sharing his expertise on the  NPRG, and for questioning all our assumptions. 

\appendix

\section{Algebraic objects}
\label{a:algebraic-objects}
The construction for a basis of the $Q$-state Potts model works as follows: 
Chose   vectors $\vec e_\alpha\in \mathbb{R}^{n}$, with 
\be
n:=Q-1\ ,
\ee   s.t.
\be\label{4}    \vec e_{\a} \cdot \vec e_\beta := 
\sum_{i}  e_\alpha^i    e_\beta^i =   \delta_{\alpha\beta} - \frac1Q 
 \ .
\ee
We use a dot for scalar products in $i$-space (roman indices, dimension $n=Q-1$, {\em unconstrained basis}); a circ ``$ \circ$''   denotes the scalar product in $\alpha$-space (greek indices, dimension $Q$, {\em constrained basis}).
These vectors can be constructed recursively, see \cite{Golner1973,ZiaWallace1975}, Eq.~(2). 
They satisfy
\be\label{5}
\sum_\alpha e_i^\alpha =  0.
\ee
Proof: 
\be\nn
\sum_i \left( \sum_\alpha e^i_\alpha \right)^{\!2} = \sum_{\alpha \beta i} e^i_\alpha e^i_\beta = \sum_{\alpha \beta} \left(   \delta _{\alpha\beta} -\frac1Q \right) =0.
\ee
The inverse relation is 
\be\label{6}
 e^i \circ e^j = \sum_{\alpha} e^i_\alpha  e^j_\alpha =  \delta ^{ij}  .
\ee
Proof: Since the $ e^j_\beta$ form an over-complete basis, we can write an arbitrary vector as $A^j = \sum_{\beta}a^\beta  e^j_\beta $; applying the tensor  in Eq.~(\ref{6}) to $A^{j}$ yields
\bea
\sum_{\alpha} e^i_\alpha  e^j_\alpha A^j &=& \sum_{\alpha j \beta }  e^i_\alpha  e^j_\alpha a^\beta  e^j_\beta = \sum_{\alpha\beta}  e^i_\alpha   a^\beta \left(\delta_{\alpha \beta}-\frac1Q\right) =   A^i\ .
\eea
To arrive at the last line we used \Eq{5}. This proves \eq{6}.

\section{The structure factor $C$ in dimensions $0$ and $1$}
\label{C-d=0+1}
It is instructive to consider 3-point properties for the two solvable cases $d=0$ and $d=1$. 
In $d=0$ there is only one cluster, so   $C$   becomes 
\be
C_{d=0}=\frac{P(\mbox {all 3 points in the same cluster})}{P(\mbox {2 points in the same cluster})^{3/2}} = 1.
\ee
We can also ask that all of them are in cluster say 1, then 
\be
\frac{P(\mbox {all 3 points in     cluster 1})}{P(\mbox {2 points in     cluster 1})^{3/2}} = \frac{\frac1Q}{(\frac1Q)^{3/2}} = \sqrt{Q}.
\ee
In $d=1$ we can order the three points (1,2,3), with distances $x=(1,2)>0$ and $y=(2,3)>0$. 
Then 
\be
\frac{\left< \Phi_1 \Phi_2 \Phi_3\right>}{\sqrt{\left< \Phi_1 \Phi_2 \right>\left<  \Phi_2 \Phi_3\right>\left< \Phi_1 \Phi_3\right>}} = 
\frac{P_{123}}{\sqrt{P_{12} P_{23} P_{13}}} =  \frac{p(x+y)}{\sqrt{p(x) p(y) p(x+y)}} =  \sqrt{\frac{p(x+y)}{ {p(x) p(y) }}}.
\ee
where $p(x)$ is the probability that two randomly chosen points at distance $x$ are in the same cluster. 
Now consider the cluster which contains point 1, and then advance to the right. The probability that the next point is still in the same cluster is $\rho\le 1$, and so on, s.t.~$p(x)=\rho^x$. This Markovian property   implies that 
\be
C_{d=1}=\frac{\left< \Phi_1 \Phi_2 \Phi_3\right>}{\sqrt{\left< \Phi_1 \Phi_2 \right>\left<  \Phi_2 \Phi_3\right>\left< \Phi_1 \Phi_3\right>}} =1.
\ee


\ifx\doi\undefined
\providecommand{\doi}[2]{\href{http://dx.doi.org/#1}{#2}}
\else
\renewcommand{\doi}[2]{\href{http://dx.doi.org/#1}{#2}}
\fi
\providecommand{\link}[2]{\href{#1}{#2}}
\providecommand{\arxiv}[1]{\href{http://arxiv.org/abs/#1}{#1}}
\providecommand{\hal}[1]{\href{https://hal.archives-ouvertes.fr/hal-#1}{hal-#1}}
\providecommand{\mrnumber}[1]{\href{https://mathscinet.ams.org/mathscinet/search/publdoc.html?pg1=MR&s1=#1&loc=fromreflist}{MR#1}}

\newpage
\tableofcontents

\end{document}